\documentclass[notitlepage,letterpaper,twocolumn,nofootinbib,superscriptaddress]{revtex4-2}
\usepackage{amssymb,amsmath,amsfonts}
\usepackage{bm}
\usepackage[usenames,dvipsnames]{color}
\usepackage[breaklinks,colorlinks,]{hyperref}
\usepackage{physics}
\usepackage{mathtools}
\usepackage{xcolor}
\usepackage{xspace}
\usepackage{enumitem}
\setitemize{itemindent=0.02in}

\definecolor{mycolor}{HTML}{377eb8} 
\hypersetup{
  linkcolor  = mycolor,
  citecolor  = mycolor,
  urlcolor   = mycolor,
  colorlinks = true
}

\newcommand{\jax}{\texttt{JAX}\xspace}
\newcommand{\galax}{\texttt{galax}\xspace}

\begin{document}

\title{A differentiable forward model for weakly perturbed stellar streams:\\
substructure forecasts from density and kinematics spectra}

\author{Noemi Anau Montel}
\email[ ]{noemiam@mpa-garching.mpg.de}
\affiliation{Max-Planck-Institut für Astrophysik, Karl-Schwarzschild-Str.\ 1, 85748 Garching, Germany}

\author{Fabian Schmidt}
\email[ ]{fabians@mpa-garching.mpg.de}
\affiliation{Max-Planck-Institut für Astrophysik, Karl-Schwarzschild-Str.\ 1, 85748 Garching, Germany}

\begin{abstract}
    Stellar streams are a promising way to gravitationally detect low-mass substructure, since their low dynamical temperature makes them retain the imprint of weak gravitational perturbations.
    We develop a fast, differentiable forward model for perturbed stellar streams in the diffusion regime, where the stream is heated by many small velocity kicks rather than by a few strong encounters. The substructure population enters only through its power spectrum, so the computational cost is insensitive to the number of perturbers, and alternative dark matter models and/or baryonic perturbers can be explored by changing this single input.   
    We validate the simulations against analytical predictions, then forecast the sensitivity of a GD-1-like stream to the substructure power spectrum, adding to the stream density the full kinematics, both proper motions and the radial velocity. 
    Kinematic information tightens the constraints by a factor of $\sim 3$--$5$ relative to density alone, improving the precision on the dark matter free-streaming cutoff scale from $\sim 1.2$~dex to $\sim 0.25$~dex at a fiducial value of $M_{\rm hm} = 10^6 M_\odot$ for a $5$~Gyr stream. A single well-measured stream could thus constrain dark matter competitively with current limits from strong lensing and satellite counts.    
\end{abstract}

\maketitle

\section{Introduction}  \label{sec:intro}

Determining the microphysical nature of dark matter, which constitutes the majority of the matter content of the Universe, remains a central challenge in modern physics. On cosmological scales (of order of comoving Mpc and above), a wide range of observations are well described by the $\Lambda$CDM paradigm, in which dark matter behaves as a cold, collisionless component governing the growth of structure \cite{Peebles_1982_CDM}. This paradigm predicts a large-scale distribution of matter and properties of massive halos that are largely in agreement with observational data. In contrast, the behavior of dark matter on small scales remains largely unconstrained, leaving open the possibility that additional microphysical properties, such as non-negligible thermal velocities \cite{Hogan_2000_WDM}, self-interactions \cite{Spergel_2000_SIDM}, or wave-like behavior \cite{Ferreira_2021_ULDM}, could manifest through subtle modifications to the abundance and internal structure of low-mass halos \cite{deMartino:2020gfi}.

The main obstacle in probing these small-scale regimes is that the lowest-mass dark matter structures are expected to host little or no luminous matter, leaving us without a direct tracer of the underlying density field, and rendering them effectively invisible to traditional astronomical surveys \cite{Belokurov_2013, Fitts_2017}. Nevertheless, their gravitational influence is irreducible and can be probed through its subtle effects on observable structures. One such precision probe of their gravitational imprint are stellar streams. Formed from the tidal disruption of globular clusters or dwarf galaxies, these streams are dynamically cold and sensitive tracers of the underlying gravitational potential \cite{Johnston_2002, Ibata_2002}. Excitingly, observations of well-studied stellar streams have consistently revealed density variations that cannot be explained by a smooth, featureless halo, hinting that dark matter substructures are leaving their gravitational imprint on these streams \cite{Bonaca:2024dgc}.

Small-scale perturbations (e.g.~from dark matter subhalos) imprint characteristic structure on stellar streams, transforming otherwise smooth, dynamically cold debris into systems with gaps, clumps, and coherent track deviations. Early studies showed that repeated encounters can dynamically heat streams and increase their dispersion, potentially erasing thin structures if substructure is too abundant \cite{Johnston_2002, Ibata_2002}. Subsequent studies emphasized that localized density perturbations produced by individual flybys can be used as a characteristic signature of substructure, which generate asymmetric features and ``gaps'' along the stream \cite{Siegal-Gaskins_2008, Carlberg_2009, Yoon_2011}, that can evolve over time \cite{Erkal_2015a}. 

Many studies have therefore focused on modeling individual subhalo encounters and their imprints on stellar streams, using features such as gaps and track deviations to infer the properties of the perturbers \cite[e.g.][]{Erkal_2015b, Bonaca_2019}, more recently also exploiting the stream kinematics through simulation-based inference \cite{2025arXiv251207960N}.
However, if a population of dark subhalos exists, stellar streams should be continuously perturbed by many such objects over their lifetime, leading to a superposition of signatures rather than isolated features. 
This motivates statistical descriptions of gap production and evolution, which show that the number and size distribution of gaps depend on both the subhalo mass function and the age and orbital history of the stream \cite{Carlberg_2013, Erkal_2016, Erkal_2017}.
To describe this more complex regime, later work has treated the stream’s one-dimensional density and track as the accumulated response to many perturbations rather than as the outcome of a single encounter. In particular, Ref.~\cite{Bovy_2016} introduced a linear perturbation framework in action-angle space that follows the effect of subhalo flybys on a cold stream and showed that the resulting density and track power spectra encode the subhalo mass function. Ref.~\cite{Banik_2019} then applied this idea to Pal 5 and showed that baryonic perturbers, especially the Galactic bar and giant molecular clouds, can contribute power comparable to that from dark subhalos and therefore must be modeled explicitly. Building on that framework, Refs.~\cite{Banik_2021a, Banik_2021b} used the power spectra of GD-1 and Pal 5 to infer the abundance of dark subhalos and to place constraints on dark matter models that suppress small-scale structure.
More recent works have both extended the modeling flexibility and developed more general analytic foundations. For example, Ref.~\cite{Nibauer:2024uue} developed a Hamiltonian perturbation framework in observable coordinates that handles time-dependent potentials and baryonic perturbers, and Ref.~\cite{Nibauer_2026} applied it to GD-1 kinematics to place the first kinematic constraints on the subhalo population in the $10^5-10^9\,\rm M_\odot$ mass range. However, resolving individual encounters becomes computationally expensive as $M_{\rm min}$ decreases, since the number of subhalo impacts grows steeply toward low masses.
An important theoretical step was taken by Ref.~\cite{Delos_2022}, who worked in the diffusion regime (where the stream accumulates many small perturbations from a continuous substructure distribution) and derived a closed-form relation between the 3D power spectrum of the dark matter density field and the 1D power spectrum of the stream density and velocity. This framework provides the theoretical foundation on which the present paper builds. 

In this work, we present a fast, differentiable forward model for perturbed stellar streams in the diffusion regime. Our approach combines the analytical velocity injection formalism of Ref.~\cite{Delos_2022} with direct numerical integration of stream orbits and explicit treatment of stream formation, providing a precise statistical description of substructure perturbations coupled to a realistic dynamical model. The main methodological difference from existing works  \cite[e.g.][]{Bovy_2016, Banik_2021a, Banik_2021b, Nibauer:2024uue, Nibauer_2026} is that we do not resolve individual substructure encounters, but we instead imprint the collective statistical effect of an entire substructure population onto the stream stars. 

This scales naturally to the low-mass regime where discrete-encounter methods become prohibitive for Bayesian inference analyses that require many forward-model evaluations, and depends on the perturbing environment only through its power spectrum $\mathcal{P}(q)$, which is straightforward to extend to non-CDM dark matter scenarios and in principle accommodate baryonic contributions. We further stress that, as in Refs.~\cite{Nibauer:2024uue, Nibauer_2026}, our model evolves the stream through direct particle integration rather than a perturbative treatment in action-angle coordinates \cite{Bovy_2016,Banik_2021a, Banik_2021b}.

As a first application, we forecast the sensitivity of a GD-1-like stream to the parameters of the substructure power spectrum. Going beyond the density power spectrum, we exploit the full set of kinematic observables (the two on-sky proper motions and the radial velocity) motivated by the new kinematic data now becoming available: Gaia proper motions across GD-1 \cite{Tavangar_2025}, recent DESI radial velocities for several hundred members \cite{Jarvis_2026}, and dedicated upcoming facilities such as the VIA Project \cite{VIA_project}.
We find that including kinematic information tightens the forecast constraints on the substructure power spectrum parameters by a factor of $\sim 3$--$5$ relative to density alone, reaching a level potentially competitive with current limits on the dark matter free-streaming cutoff scale from strong lensing and Milky Way satellite counts \cite[e.g.][]{Enzi_2021, Nadler_2021}.

The paper is organized as follows. In Sec.~\ref{sec:diffusion}, we review the analytical description of stellar streams in the diffusion regime. In Sec.~\ref{sec:simulations}, we describe our forward model, and validate the velocity injection implementation against the analytical predictions in both an idealized setup and a realistic one including orbital dynamics and stream formation. In Sec.~\ref{sec:forecasts}, we apply the forward model to forecast the sensitivity of a GD-1-like stream to substructure parameters, using both the stream density and the full kinematic power spectra. We discuss the limitations of the framework, comparisons to previous works, and future directions in Sec.~\ref{sec:discussion}, and conclude in Sec.~\ref{sec:conclusion}.

\section{Stellar Streams in the Diffusion Regime} \label{sec:diffusion}

In this section, we briefly review the analytic perturbative description of stellar streams in the diffusion regime, where stars are subjected to many small velocity kicks due to substructure encounters, rather than isolated large encounters. In this limit, stream stars perform a random walk around in phase space, which can be understood at the analytical level. We follow Ref.~\cite{Delos_2022}, who first derived the analytical connection between the statistic of the substructure density field perturbing the stars and the stellar stream statistics. 

\subsection{Velocity injection formalism} \label{subsec:kicks}

\begin{figure}
    \centering
    \includegraphics[width=0.75\linewidth]{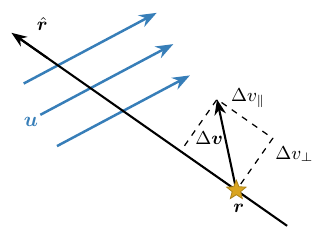}
    \caption{Geometry of the velocity injection formalism, adapted from Ref.~\cite[Fig.~1]{Delos_2022}. The stellar stream (black line) runs along $\hat{\boldsymbol{r}}$, and the blue arrows represent the relative velocity field $\boldsymbol{u}$ of the dark matter substructure environment. A star at position $\boldsymbol{r}$ receives a velocity impulse $\Delta\boldsymbol{v}$, decomposed into a component $\Delta v_{\parallel}$ parallel to the stream and a component $\Delta\boldsymbol{v}_{\perp}$ transverse to it.}
    \label{fig:geometry}
\end{figure}

Assuming the self-gravity of a stellar stream's star is negligible and that each star can be considered independently, a relationship between the substructure environment and the integrated velocity change $\Delta\boldsymbol{v}$ that it injects on to the star can be found. Consider a star moving with velocity $\boldsymbol{v}$ through a substructure environment with density in Fourier space $\rho(\boldsymbol{q}) = \bar{\rho}[1 + \delta(\boldsymbol{q})]$, where $\bar{\rho}$ is the mean density and $\delta(\boldsymbol{q})$ represents density fluctuations (see Fig.~\ref{fig:geometry} for the setup). With relative velocity $\boldsymbol{u}$ between the star and substructure, the integrated velocity change over time $t$ of a star at position $\boldsymbol{r}$ is
\begin{equation}\label{eq:kick}
    \Delta\boldsymbol{v}(\boldsymbol{r}) = \int\frac{\mathrm{d}^3\boldsymbol{q}}{(2\pi)^3}\, \delta(\boldsymbol{q})\, \mathrm{e}^{i\boldsymbol{q}\cdot\boldsymbol{r}}\, \boldsymbol{V}^*(\boldsymbol{q}|\boldsymbol{u},t),
\end{equation}
where the velocity kernel is
\begin{equation} \label{eq:response}
    \boldsymbol{V}(\boldsymbol{q}|\boldsymbol{u},t) = 8\pi i G\bar{\rho}\, \mathrm{e}^{i\boldsymbol{q}\cdot\boldsymbol{u} t/2} \frac{\sin(\boldsymbol{q}\cdot\boldsymbol{u} t/2)}{\boldsymbol{q}\cdot\boldsymbol{u}} \frac{\boldsymbol{q}}{q^2}.
\end{equation} 

Here $q \equiv |\boldsymbol{q}|$ denotes the magnitude of the wavenumber vector. The kernel $\boldsymbol{V}(\boldsymbol{q}|\boldsymbol{u},t)$ arises from integrating the Fourier space gravitational acceleration of the star due to the density fluctuations $\delta(\boldsymbol{q})$, i.e.
\begin{align}
  \Delta\boldsymbol{v}(\boldsymbol{r}) &= \int_0^t dt' \dot{\boldsymbol{v}}(\boldsymbol{r},t') \nonumber\\
  &= -4\pi i G\bar{\rho}\int_0^t dt' \int \frac{\mathrm{d}^3\boldsymbol{q}}{(2\pi)^3}\,\delta(\boldsymbol{q})\,\mathrm{e}^{i\boldsymbol{q}\cdot[\boldsymbol{r}+\boldsymbol{u} t']}\,\boldsymbol{q}/q^2 \; ,
\end{align} 
over the time interval $[0,t]$, during which the star drifts through the substructure field at relative velocity $\boldsymbol{u}$. This drift introduces a time-dependent phase $\mathrm{e}^{-i\boldsymbol{q}\cdot\boldsymbol{u} t'}$, whose integration over $t' \in [0,t]$ produces the $\sin(\boldsymbol{q}\cdot\boldsymbol{u}\,t/2)/(\boldsymbol{q}\cdot\boldsymbol{u})$ factor in Eq.~\eqref{eq:response}.
The kernel thus encodes the coherent accumulation of the gravitational pull as the substructure drifts past the star over the interval $t$. In the simulation (Sec.~\ref{subsec:sim_diffusion}), $t$ corresponds to the kick interval $\Delta t$, so that the cumulative effect of the substructure environment is built up as a sequence of discrete kicks, each computed using $\boldsymbol{V}(\boldsymbol{q}|\boldsymbol{u},\Delta t)$.

The assumptions needed to derive this result are that the substructure environment is not perturbed by the presence of the star, and that the velocity injection is small compared to the star's velocity, $\Delta \boldsymbol{v} \ll \boldsymbol{v}$, so that the relative trajectory is not significantly perturbed by the velocity injection. Both assumptions are accurate in the regime we are interested in here (in fact, they are usually assumed also when considering massive perturbers).

This formalism easily extends to a substructure velocity distribution $f(\boldsymbol{u})\mathrm{d}^3\boldsymbol{u}$ by decomposing the density field into independent components $\delta(\boldsymbol{q}) = \sum_{i=1}^N \delta_i(\boldsymbol{q})$, each moving with velocity $\boldsymbol{u}_i$ sampled from $f(\boldsymbol{u})$.\footnote{This assumes all subfields share the same power spectrum and are statistically independent, neglecting correlations between density structures at different velocities.} In the continuum limit, summation over components becomes integration over velocities.

The observationally relevant regime occurs when $qut \gg 1$, i.e. the relative distance traversed exceeds the substructure scale, since stellar stream ages ($\sim\text{Gyr}$) far exceed typical encounter timescales $(qu)^{-1} \sim \text{Myr}$. This defines the diffusion regime: velocity kicks accumulate as a random walk, and only density modes perpendicular to the relative velocity contribute (those with $\boldsymbol{q} \cdot \boldsymbol{u} = 0$).

\subsection{Connection between dark matter environment and stellar stream statistics}

Given the velocity injection setup, it is possible to derive the one-dimensional power spectrum of velocity injections along the stream, that follows from the correlation function of velocity kicks at separated points. Assuming isotropic Maxwellian substructure velocities with dispersion $u_0$ in the Galactic frame, Ref.~\cite{Delos_2022} derived
\begin{equation} \label{eq:pdvpar}
\begin{split}
    P_{\Delta v,\parallel}(k,t) &= 16\pi^4 G^2\bar{\rho}^2 \frac{\sqrt{2/\pi}}{u_0} k^2 t \\
    &\quad \times \int_k^\infty \frac{\mathrm{d}q}{q} \frac{\mathcal{P}(q)}{q^6} \exp\!\left(-\frac{1}{2}\frac{k^2}{q^2}\frac{v^2}{u_0^2}\right),
\end{split}
\end{equation}
where $\mathcal{P}(q) \equiv [q^3/(2\pi^2)]P^\text{sub}(q)$ is the dimensionless substructure power spectrum, $k$ is the wavenumber along the stream, and $v$ is the stream's velocity. The integral restricts to $q > k$: only substructure modes finer than the length scale $k^{-1}$ along the stream contribute. The exponential suppresses contributions when stream velocity exceeds substructure velocities, as rapid relative motion reduces kick efficiency.

The stream's phase-space response to velocity kicks due to many small encounters can be treated via the Fokker--Planck equation. In the diffusion regime, individual encounter displacements are small compared to the width of the stream's phase-space distribution, justifying a treatment where the distribution function $f(x,v,t) = f_0(v,t) + f_1(x,v,t)$ is decomposed into a spatially uniform background component $f_0$ and a small spatially varying perturbation $f_1$. The derivation assumes the stellar stream is an unbound one-dimensional system with uniform zeroth-order density. Here, $f_0$ is taken to be Maxwellian with a velocity dispersion $\sigma^2(t)$ that broadens over time due to the cumulative heating from encounters, while $f_1$ is treated at linear order and captures all perturbations to the stream density and velocity.
By integrating the expression for $f_1(x,v,t)$ over velocities and dividing by the unperturbed stream density $\bar{\rho}_*$ to obtain the stream density contrast $\delta^*(k,t)$, Ref.~\cite{Delos_2022} derived the stream density power spectrum
\begin{equation} \label{eq:pdensity}
    P^*(k,t) = \chi_*\!\left(k\sigma_0 t, \frac{D}{k\sigma_0^3}\right) \frac{k^2 t^2}{3} P_{\Delta v,\parallel}(k,t),
\end{equation}
where $\sigma_0$ is the stream velocity dispersion, $D = \int_{k_{\rm min}}^\infty (\mathrm{d}k/\pi)\, \mathrm{d}P_{\Delta v,\parallel}/\mathrm{d}t$ is the diffusion coefficient, and $\chi_*$ is a transfer function encoding velocity dispersion damping (see Ref.~\cite[Eq.~(45)]{Delos_2022}). This derivation 
neglects spatial and temporal variations in the diffusion coefficient, and treats velocity kicks at different times as uncorrelated.

The stream velocity power spectrum follows similarly:
\begin{equation} \label{eq:pvelocity}
    P^*_{v,\parallel}(k,t) = \chi_v\!\left(k\sigma_0 t, \frac{D}{k\sigma_0^3}\right) P_{\Delta v,\parallel}(k,t),
\end{equation}
with a different transfer function $\chi_v$ (see Ref.~\cite[Eq.~(F5)]{Delos_2022}). 

\subsection{Orbital dynamics treatment} \label{subsec:orbit}

The derivation of stream perturbations above treats the stream as a one-dimensional system and neglects orbital complexities. To fully describe the stellar stream's behavior, it is necessary to account for the effects of orbital dynamics, which complicate the connection between velocity kicks $\Delta\boldsymbol{v}$ and position perturbations $\delta \boldsymbol{x}$. Ref.~\cite{Delos_2022} adopts a simplified approach by averaging the perturbations over many orbital periods and treating the orbital dynamics as a random variable.

Three key effects arise. First, velocity kicks parallel to the orbital velocity induce secular position perturbations with efficiency
\begin{equation} \label{eq:lambda}
    \delta x_\parallel = \Lambda\, \Delta v_\parallel\, t,
\end{equation}
where $\Lambda < 0$ (the ``donkey effect'': higher-energy orbits have longer periods, so energy injection causes stars to lag). The orbit-averaged factor $\langle\Lambda^2\rangle$ scales all velocity injection power spectra. 
Second, orbital expansion and contraction require a co-moving frame where separations scale as $\omega \equiv v_{\rm orb}/\langle v_{\rm orb}\rangle$. The velocity injection power spectrum in this frame becomes
\begin{equation} \label{eq:omega}
    P_{\Delta v}^{\rm co}(k) = \int \mathrm{d}\omega\, \frac{f_{\rm orb}(\omega)}{\omega}\, P_{\Delta v}(k/\omega)\Big|_{v=\omega\langle v_{\rm orb}\rangle},
\end{equation}
where $f_{\rm orb}(\omega)$ is the distribution of $\omega$ over an orbit. The physical-space stream power spectrum is then $P^*(k) = P^{*,\rm co}(\omega_0 k)$, where $\omega_0$ is the value at observation time (this transformation also applies to the velocity power spectrum $P^*_{v,\parallel}(k)$). 
Lastly, the stream's age varies along its length due to continuous sourcing, leading to a position-dependent age that affects both the density and velocity power spectra. Different parts of the stream have been perturbed for different amounts of time, so we must average the power spectra over time. This can be done by integrating the power spectra over the age of the stream, expressed as: \begin{align} P^*(k,t_\mathrm{age})=\frac{1}{t_\mathrm{age}}\int_0^{t_\mathrm{age}}\mathrm{d}t\, P^*(k,t), \end{align} which applies similarly to the velocity power spectrum $P^*_{v,\parallel}(k)$.

\subsection{Limitations of the analytical treatment}

At this point, it is worth recapping the assumptions and limitations in the different ingredients of the semi-analytical model of Ref.~\cite{Delos_2022}. 
The velocity injection formalism employed is exact in the limit of many small, temporally uncorrelated perturbations: given the substructure power spectrum $\mathcal{P}(q)$ and velocity distribution $f(\boldsymbol{u})$, the velocity-injection power spectrum $P_{\Delta v,\parallel}(k,t)$ follows without further approximation. However, several additional assumptions are needed to connect velocity injections to observable stream properties, and these introduce varying degrees of approximation:
\begin{itemize}[leftmargin=0.2in]
    \item \textit{Stream geometry.} The stream is treated as a straight, one-dimensional system with uniform zeroth-order density. Real streams follow curved orbits and exhibit density gradients along their length, both of which can affect the mapping between velocity kicks and density perturbations.
    \item \textit{Diffusion coefficient.} The diffusion coefficient $D$ is assumed to be spatially uniform, whereas in reality its spatial variation cannot be neglected. In addition, there is some ambiguity in defining which length scales of perturbations contribute to the diffusion coefficient (see Ref.~\cite[][Sec.~4.2]{Delos_2022} for a discussion).
    \item \textit{Orbital dynamics.} As discussed above, the analytic treatment averages perturbations over many orbital periods, treating the kick efficiency $\Lambda$ and the co-scaling factor $\omega$ as random variables. In reality, the long-term displacement along the stream produced by a given velocity kick depends on the orbital phase at which it occurs, introducing correlations that are not captured by orbit-averaged quantities.
    \item \textit{Stream formation.} The continuous sourcing of the stream and the resulting position-dependent age are only accounted for through a time average of the power spectra. Effects such as episodic stripping near pericenter \citep{Kuepper_2010, Kuepper_2012} are neglected.
    \item \textit{Substructure environment.} The density fields at different kick times are assumed to be statistically independent, and the substructure velocity distribution is taken to be an isotropic Maxwellian. In reality, substructure fields are temporally correlated, and the velocity distribution can be anisotropic or non-Maxwellian.
\end{itemize}
A fully self-consistent treatment that incorporates all of these effects has not been achieved analytically, and is likely to be impossible. This motivates a simulation-based approach that couples the velocity injection formalism, which accurately captures the statistical effect of the substructure environment, with a numerical treatment of orbital dynamics and stream formation.

\section{Simulations of stellar streams in the diffusion regime} \label{sec:simulations}

Building on the motivation highlighted at the end of the previous section, we develop a forward model that retains the velocity injection formalism of Sec.~\ref{subsec:kicks}, but replaces the approximate analytic treatment of stream formation and dynamics with direct numerical orbit integration, improving upon many of the aforementioned approximations. 

Our approach is fundamentally different from frameworks that resolve individual subhalo encounters \cite[e.g.][]{Bovy_2016, Nibauer_2026}. Rather than sampling and simulating discrete flybys, we model the collective statistical effect of an entire substructure population in a mass range $[M_{\rm min}, M_{\rm max}]$ on stream observables. This is further motivated by two considerations. First, simulating encounters individually rapidly becomes computationally intensive as $M_{\rm min}$ decreases, since the number of subhalos (hence encounters) grows steeply towards low masses (above the free-streaming cutoff of the dark matter, that is). Additionally, the density perturbations arising in non-standard dark matter scenarios, such as fuzzy dark matter, cannot be accurately described as discrete structures. 
Instead, we start from a density contrast field $\delta(\boldsymbol{q})$ describing the substructure environment, and map its effect directly onto small velocity kicks imparted to stream stars (following the velocity injection formalism in Sec.~\ref{subsec:kicks}), working in the diffusion regime where these kicks accumulate incoherently.

\subsection{Implementation}  \label{subsec:simulations}

We implement simulations of stellar streams in the diffusion regime in the python library \jax \cite{jax2018github}, which has seen increasing adoption in recent stellar stream analyses \cite[e.g.][]{Nibauer:2024uue, Alvey:2023pkx}. 
\jax provides three key features that motivate our choice. First, just-in-time (\texttt{jit}) compilation via XLA allows the simulation loop to be compiled once and executed efficiently on both CPUs and GPUs. Second, evolution of individual star particles is  parallel\footnote{Under the general assumption that each star can be considered independently.}, and \jax's \texttt{vmap} transformation allows this to be easily executed on GPU without any modification to the single-particle code. Third, \jax natively supports automatic differentiation (AD), providing exact derivatives of any user-defined function up to numerical precision. This is a key ingredient for gradient-based inference, for computing Fisher matrices, and for optimal data compression.

Our modelling consists of three components: initial conditions for the ejected star particles (Sec.~\ref{subsec:init}), dynamical evolution of the ejected particles under a smooth gravitational potential $\Phi$, and velocity kicks from dark matter substructure perturbations (Sections \ref{subsec:substructure} and \ref{subsec:sim_diffusion}). The first two are standard ingredients of stream simulation codes based on particle-spray techniques. In particular, for the gravitational evolution we use the open-source galactic dynamics library \galax \cite{galax}, which solves the equations of motion $\ddot{\boldsymbol{x}} = -\nabla\Phi(\boldsymbol{x})$ for each star particle via \texttt{diffrax} \cite{diffrax}, and is fully compatible with the \jax ecosystem. The novel contribution of this work is the third component: rather than simulating discrete encounters, we imprint the cumulative effect of a substructure population directly as ongoing velocity perturbations on the stream stars interlaced with the orbit integration.

\subsubsection{Initial conditions}  \label{subsec:init}

Setting the initial conditions of star particles requires specifying two distributions: the times at which stars are stripped from the progenitor $p(t_{\rm strip})$, and the phase-space distribution of each star at the moment of stripping $p(\boldsymbol{x}_{\rm init}, \boldsymbol{v}_{\rm init})$.

\paragraph*{Stripping times.}
For the stripping times, in this work we adopt uniform time intervals rather than a stripping rate tied to the progenitor's mass-loss history. In reality, tidal stripping is strongest near pericenter, where the host tidal field is most intense, leading to bursts of star ejection correlated with the orbital phase. This episodic stripping introduces density variations along the stream, e.g.~epicyclic overdensities \cite{Kuepper_2010, Kuepper_2012}, which are uncorrelated with substructure perturbations, but can complicate the interpretation of the stream power spectrum \cite{Bovy_2016}. Since subhalo impacts predominantly affect the oldest parts of the stream while episodic stripping imprints mostly on the youngest parts near the progenitor, uniform stripping is a good approximation for our purposes. More complex, physically motivated stripping rates derived from the progenitor's mass-loss history can be straightforwardly incorporated within our framework.

\paragraph*{Ejection model.}
For the phase-space distribution at stripping time, we follow a Lagrange-point stripping approach \cite[e.g.][]{Bowden_2015, Alvey:2023pkx}. Each star is assigned to the leading or trailing arm with equal probability, and placed at $\boldsymbol{x}^{\rm trail,\,lead}_{\rm init} = (1 \pm r_t/r_p)\,\boldsymbol{x}_p$, where $\boldsymbol{x}_p$ is the progenitor position at stripping time and $r_p = |\boldsymbol{x}_p|$. The tidal radius is approximated as $r_t = \sigma_p/(\sqrt{3}\omega_p)$, where $\omega_p = |\boldsymbol{x}_p \times \boldsymbol{v}_p|/r_p^2$ is the instantaneous orbital angular frequency. Since we do not follow the progenitor mass evolution, we treat the velocity dispersion $\sigma_p$ as a fixed control parameter, chosen to produce streams of sufficient length. The initial velocity of each star is set to the progenitor velocity plus a co-rotation correction at the Lagrange point, with an additional isotropic Gaussian perturbation of dispersion $\sigma_p$. More generally, any distribution function from which initial conditions can be sampled, including more sophisticated particle spray methods calibrated against $N$-body simulations \cite{Fardal_2015, Chen_2025}, may be substituted into our framework without modification.

\subsubsection{Substructure model} \label{subsec:substructure}

The effect of a dark matter substructure environment on stellar streams can be characterised by two ingredients: the power spectrum of density fluctuations $\mathcal{P}(q)$, which encodes the spatial distribution of substructure, and the velocity distribution function $f(\boldsymbol{u})$ of the substructure in the stellar stream frame.

\paragraph*{Substructure power spectrum.}
The power spectrum $\mathcal{P}(q)$ can be connected to the traditional halo model of structure formation. Assuming halo positions are uncorrelated and each halo of mass $M$ has a density profile $\rho(r|M) = M R^{-3} p(r/R)$, where $p$ is a dimensionless universal profile and $R = R(M)$ is the scale radius, the substructure power spectrum is \cite{Scherrer_1991_DM_stat} 
\begin{equation} \label{eq:Pksub}
    \mathcal{P}(q) = \frac{1}{\bar{\rho}^2} \int \mathrm{d}M\, \frac{\mathrm{d}n}{\mathrm{d}M} \left[M\tilde{p}(qR)\right]^2,
\end{equation} 
where $\bar{\rho}$ is the mean substructure density, $\mathrm{d}n/\mathrm{d}M$ is the halo mass function, and $\tilde{p}(x)$ is the 3D Fourier transform of the universal profile. The amplitude, shape, and slope of $\mathcal{P}(q)$ encode information about the abundance, mass function, and density profiles of subhalos, and can be directly connected to predictions from different dark matter theories.

It is important to note that the power spectrum is a more general description than the halo model. Density inhomogeneities that cannot be described as a collection of discrete subhalos, e.g.~the interference patterns associated with fuzzy dark matter on scales of the de Broglie wavelength \cite[e.g.][]{Ferreira_2021_ULDM}, can in principle be characterised directly through $\mathcal{P}(q)$, measured from simulations, without requiring a subhalo prescription.

\paragraph*{Substructure velocity distribution.}
The second ingredient is the velocity distribution of substructure, $f(\boldsymbol{u})$. We adopt an isotropic Maxwellian with scale velocity $u_0$ in the Galactic frame, consistently with Ref.~\cite{Delos_2022}. More realistic distributions, including anisotropic or multi-component models \cite[e.g.][]{Mao_2013, zhang2026setnightfirebuilding}, can be straightforwardly substituted as long as they can be sampled.

\subsubsection{Velocity kicks implementation}  \label{subsec:sim_diffusion}

\begin{figure*}[t]
    \centering
    \includegraphics[width=\linewidth]{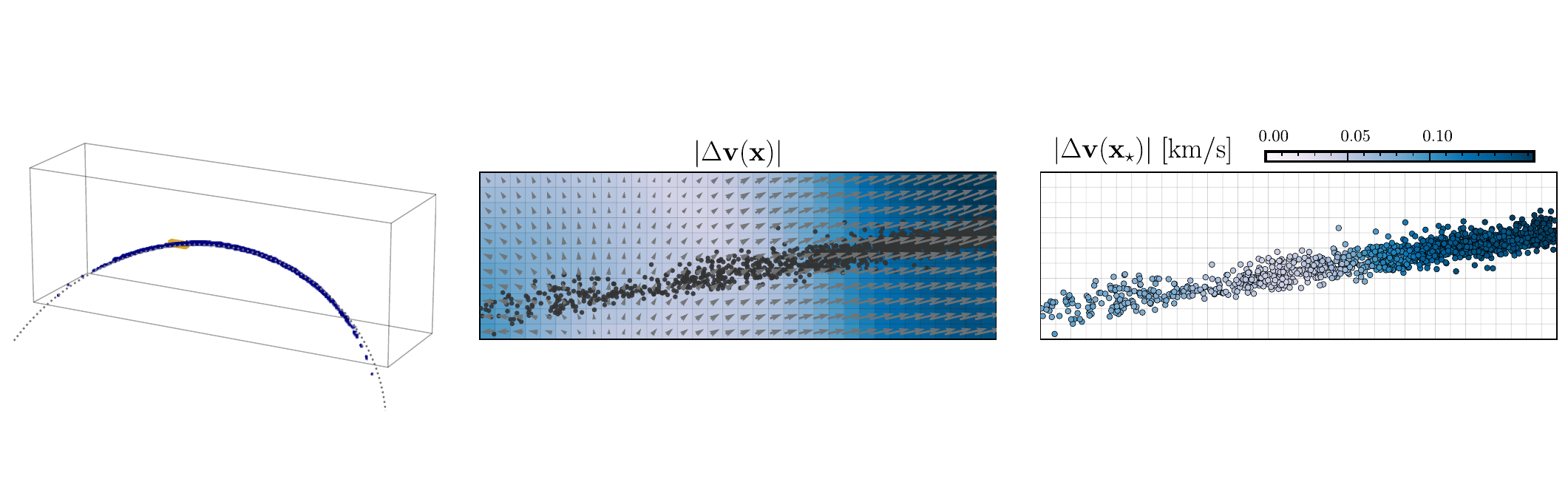}
    \vspace{-1.cm}
    \caption{Illustration of the velocity kick implementation at a single kick time $t_k$, from an actual simulation realization. \textit{(Left:)} The stream (blue points) within the 3D grid used to discretize the substructure density field. The grid is aligned with the instantaneous stream velocity direction and sized to encompass the stream extent with additional padding. The small golden box indicates the region shown in the middle and right panels. \textit{(Center:)} The velocity kick field $\Delta\boldsymbol{v}(\boldsymbol{x})$, defined in Eq.~\eqref{eq:kick}, within the highlighted region, averaged over its $z$-depth. The color encodes the kick amplitude $|\Delta\boldsymbol{v}|$ and the arrows indicate the kick direction at each grid node at the middle of the $z$-slab. Stream stars are overlaid as points. The resolution of the grid is $0.05$ kpc. \textit{(Right:)} The kick field evaluated at the irregular star positions via NUFFT, yielding an individual velocity kick $\Delta\boldsymbol{v}(\boldsymbol{x}_\star)$ for each star. Marker colors reflect the kick amplitude.}
    \label{fig:grid}
\end{figure*}

Having specified the substructure model, we now describe how velocity kicks due to the substructures are computed and applied to the stream particles. At each of $N_k$ discrete kick times, we sample a realisation of the substructure density field, compute the resulting velocity kick field in Fourier space via Eq.~\eqref{eq:response}, and evaluate the kicks at the star positions. The following paragraphs describe each step in detail.

\paragraph*{Grid setup.} 
The subtructure density field is discretised on a 3D slab with grid dimensions $(N_x, N_y, N_z)$ and physical size $(L_x, L_y, L_z)$, aligned with the instantaneous stream velocity direction so that $\hat{x}$ coincides with the stream axis (see Fig.~\ref{fig:grid}). This minimises the required longitudinal extent and concentrates resolution where it is needed. The box physical dimensions are set once from the final unperturbed stream extent, with a fractional buffer to accommodate perturbations, and held fixed throughout the simulation. The grid resolution $\eta \equiv L_i/N_i$ with $i\in\{x, y, z\} $ is the same along all three axes and is a free parameter. The sizes $N_x$, $N_y$, $N_z$ are chosen to be 5-smooth numbers (products of powers of 2, 3 and 5 only) to ensure FFT efficiency.

\paragraph*{Kick times.} 
The key free parameter controlling the temporal discretization is the kick interval $\Delta t \equiv t_k - t_{k-1}$, with $k = 1, \ldots, N_k$, which sets the duration of the integration window over which each substructure field acts on the stars, entering directly as the time argument in the velocity response kernel in Eq.~\eqref{eq:response}. Given $\Delta t$ and the total stream age $t_{\rm age}$, the kick times are uniformly spaced over the stream lifetime, with the total number of kicks $N_k$ determined accordingly.

\paragraph*{Substructure environment sampling.} 
At each kick time $t_k$, we sample $N_\delta$ independent Gaussian random fields $\delta_i(\boldsymbol{q}) \sim \mathcal{N}(0, \mathcal{P}(q))$ in Fourier space, each associated with an independent substructure velocity in the stream frame, i.e. ${\boldsymbol{u}}_i = \tilde{\boldsymbol{u}}_i - \boldsymbol{v}_{\rm stream}$, where $\tilde{\boldsymbol{u}}_i \sim f(\tilde{\boldsymbol{u}})$ is drawn from the substructure velocity distribution in the Galactic frame and $\boldsymbol{v}_{\rm stream}$ is the mean stream velocity at $t_k$. In this setup, fields are independent across both subfields and kick times.

\paragraph*{Velocity kick computation.} 
The velocity kick field in Fourier space is obtained, following the velocity injection formalism (Sec.~\ref{subsec:kicks}), by multiplying each density component by its response kernel and summing, 
\begin{equation}     
    \Delta\boldsymbol{v}(\boldsymbol{q}) = \sum_{i=1}^{N_\delta} \delta_i(\boldsymbol{q})\, \boldsymbol{V}^*(\boldsymbol{q}|\boldsymbol{u}_i, \Delta t). 
\end{equation} 
To evaluate $\Delta\boldsymbol{v}$ at the irregular star positions $\{\boldsymbol{x}_\star\}$ from the Fourier-space grid, we use a type-2 non-uniform Fast Fourier Transform (NUFFT) implemented via \texttt{jax-finufft}\footnote{{\url{https://github.com/flatironinstitute/jax-finufft}}}, which avoids interpolation onto a regular grid and evaluates Eq.~\eqref{eq:kick} directly at the star positions. The resulting kicks are rotated back to the Galactic frame and added to each star's velocity. Fig.~\ref{fig:grid} illustrates this procedure for a single kick time, using the realistic simulation setup of Sec.~\ref{subsubsec:realistic} with number of stars $N_\star=26000$, showing the stream within the 3D grid and a zoom into a subregion displaying the velocity kick field and the kicks evaluated at the star positions. In practice, a small fraction of stars can fall outside the grid boundaries (and not be affected by the kicks), either because substructure-induced perturbations in the high-$M_{\rm max}$ scenarios have displaced them too much or because the curvature of the orbit causes the stream to extend beyond the box aligned with the mean velocity direction. We monitor these occurrences throughout the simulation; in the setup considered here, around $\sim 0-5\%$ of stars are affected, depending on the $M_{\rm max}$ scenarios and initial conditions, and these are the outermost stars that lie in the low-density tails excluded from the analysis.

\paragraph*{Simulation loop.} 
The full simulation is implemented as a \texttt{jax.lax.scan} over $N_k$ kick times, where each step consists of two sub-steps: (i) orbital integration of all star particles from the previous kick time to the current  one via \texttt{galax}, and (ii) application of the velocity kick. Within each step, per-star computations are parallelised via \texttt{vmap}. Stars that have not yet been stripped at a given kick time are carried along as inactive placeholders, i.e.~they receive zero kicks and are integrated from their stripping time rather than the step start, but are retained in the  arrays to keep shapes static and allow \texttt{jit} compilation. After the final kick, all stars are integrated forward to the observation time $t = 0 \rm{\ Myr}$.

For \texttt{jit} compilation, both the kick times and the stripping times must be fixed at compile time, since their ordering determines the control flow (which stars are active at which kick). Throughout this work we assume uniformly spaced stripping and kick times, which satisfies this requirement. If stripping times were instead drawn based on a realistic progenitor cluster mass-loss rate, their ordering relative to the kick times would change across realisations, preventing \texttt{jit} compilation without recompilation for each new draw. We discuss computational cost and scaling in Appendix~\ref{app:runtime}.

Unless stated otherwise, we adopt $N_\delta=1$, $\Delta t = 90$~Myr, and $\eta = 0.05$~kpc throughout, and test convergence with respect to all three parameters in Appendix~\ref{app:convergence}.

\subsection{Validation with analytical description} \label{subsec:comparison}

To validate our simulation framework, we compare the stream density and velocity power spectra against the analytical predictions of Ref.~\cite{Delos_2022}. We proceed in two steps. In Sec.~\ref{subsubsec:idealized}, we test the velocity injection implementation in isolation by applying 3D diffusive kicks to a flat, one-dimensional stream of uniformly spaced stars with no orbital dynamics, matching the idealized setup of Ref.~\cite{Delos_2022}. Second, in Sec.~\ref{subsubsec:realistic}, we run full realistic simulations with orbital integration and stream formation.

Both setups share the same substructure environment. The dark matter subhalo mass function follows $\mathrm{d}n/\mathrm{d}M \propto M^{-2}$, normalised to a number density of $5.86 \times 10^{-4}$~kpc$^{-3}$ in the mass range $[10^6, 10^7]\,\mathrm{M}_\odot$. Subhalos are modelled with Hernquist profiles with scale radius $R = 1.05\,(M/10^8\,\mathrm{M}_\odot)^{1/2}$~kpc, and their velocity distribution is Maxwellian with scale velocity $u_0 = 120$~km\,s$^{-1}$.

\subsubsection{Idealized simulations} \label{subsubsec:idealized}

\begin{figure*}[t]
    \centering
    \includegraphics[width=\linewidth]{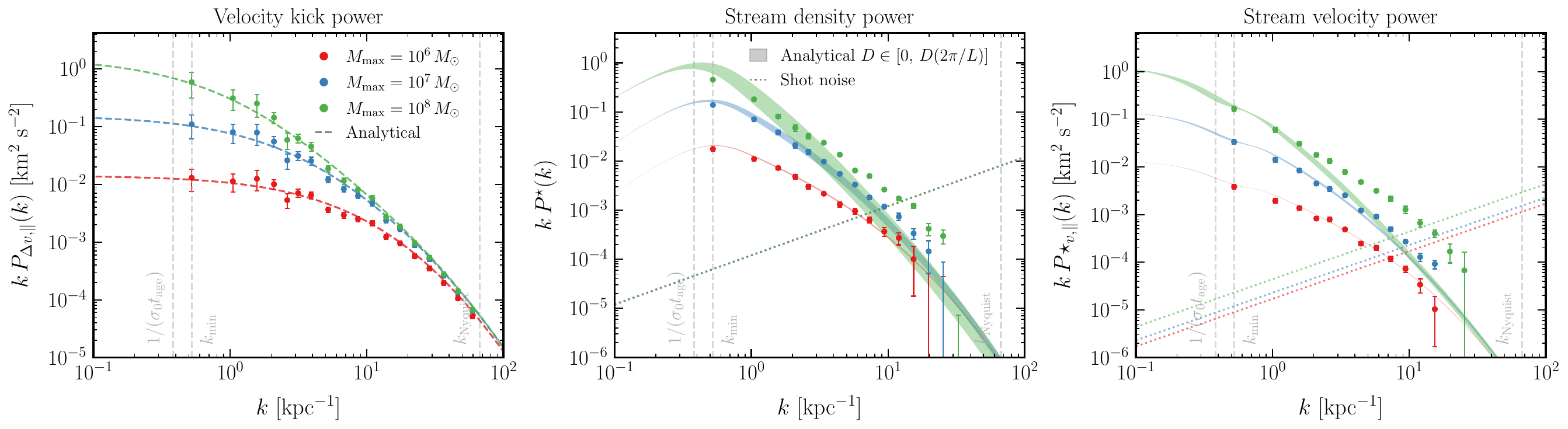}
    \caption{Validation of the simulation framework against analytical predictions in the idealized (non-orbital) setup. A uniform stream of $N_* = 10^5$ stars spanning $L = 12$~kpc is evolved for $t_{\rm age} = 7$~Gyr under diffusive velocity kicks, with initial velocity dispersion $\sigma_0 = 0.365$~km\,s$^{-1}$ and zero stream velocity. Three substructure scenarios are shown, with $M_{\rm min} = 10^5\,\mathrm{M}_\odot$ and $M_{\rm max} = 10^6$ (red), $10^7$ (blue), and $10^8\,\mathrm{M}_\odot$ (green). All power spectra are evaluated at the final time $t_{\rm age}$. \textit{Left:} velocity-injection power spectrum $k P_{\Delta v,\parallel}(k)$. \textit{Center:} dimensionless stream density power spectrum $k P^*(k)$. \textit{Right:} stream velocity power spectrum $k P^*_{v,\parallel}(k)$. Points show the mean over 100 realizations; error bars indicate the standard deviation of the mean. The shot noise floor has been subtracted from the simulation results. Dashed curves show the analytical predictions (Eq.~\eqref{eq:pdvpar} for the left panel); shaded bands in the center and right panels bracket the predictions for the density and velocity power spectra (Eqs.~\eqref{eq:pdensity} and \eqref{eq:pvelocity}) with $D = 0$ and $D = D(2\pi/L)$ (see text). Dotted lines indicate the shot noise level. Vertical dashed lines mark the minimum wavenumber $k_{\rm min} = 2\pi/L$, the free-streaming scale $1/(\sigma_0 t_{\rm age})$, and the Nyquist frequency $k_{\rm Nyquist}$.}
    \label{fig:idealized}
\end{figure*}

The stream is initialized as a uniform line of $N_* = 10^5$ stars spanning $L = 12$~kpc, with initial velocity dispersion $\sigma_p = 0.365$~km\,s$^{-1}$ and zero stream velocity ($v = 0$~km\,s$^{-1}$). Stars are evolved for $t_{\rm age} = 7$~Gyr under the diffusive kicks described in Sec.~\ref{subsec:sim_diffusion}, with no gravitational potential. We consider three substructure scenarios with fixed minimum mass $M_{\rm min} = 10^5\,\mathrm{M}_\odot$ and varying maximum mass $M_{\rm max} \in \{10^6, 10^7, 10^8\}\,\mathrm{M}_\odot$.

Fig.~\ref{fig:idealized} shows the comparison between simulations and the analytical predictions of Sec.~\ref{sec:diffusion}. The left panel shows the velocity-injection power spectrum $P_{\Delta v,\parallel}(k)$, Eq.~\eqref{eq:pdvpar}, the center panel the stream density power spectrum $P^\star(k)$, Eq.~\eqref{eq:pdensity}, and the right panel the stream velocity power spectrum $P^\star_{v,\parallel}(k)$, Eq.~\eqref{eq:pvelocity}. 
For each scenario, we run $100$ realizations and report the mean power spectra, with error bars indicating the standard deviation of the mean. The estimated Poisson noise floor, $P_{\rm shot} = L/N_\star$ for the density and $P_{\rm shot}^v = \sigma_0^2 L/N_\star$ for the velocity (with $\sigma_0$ the velocity dispersion of stream stars today), is subtracted from the simulation power spectra. 
For the density and velocity power spectra, the shaded bands bracket the analytical predictions computed with $D=0$ and $D=D(2\pi/L)$. As discussed in Ref.~\cite[][Sec.~4.2]{Delos_2022}, the diffusion coefficient $D$ is in principle spatially varying, but no analytic solution exists for this case. Setting $D=0$ neglects the induced velocity dispersion entirely, while  $D=D(2\pi/L)$ includes contributions from velocity injections up to the scale of the stream, which tends to overestimate the local velocity dispersion at small scales and hence underestimate the stream power. The true solution is expected to lie between these two limiting cases, and indeed the simulation results consistently fall within this bracket across all three substructure scenarios. 
The vertical dashed line at $k = 1/(\sigma_0 t_{\rm age})$ marks the free-streaming scale, which separates two regimes: at large scales $k \ll 1/(\sigma_0 t_{\rm age})$, the velocity dispersion has negligible effect and the density power spectrum grows as $P^\star(t) \propto t^3$, while at small scales $k \gg 1/(\sigma_0 t_{\rm age})$, streaming of stars suppresses power and the spectrum enters a steady state where continuous injection of new power balances the velocity dispersion-induced damping.

For the $M_{\rm max} = 10^8\,\mathrm{M}_\odot$ scenario, the simulation points exceed the analytical bracket at $k \gtrsim 5$~kpc$^{-1}$. The parameter distinguishing the three scenarios is the heating ratio $Dt_{\rm age}/\sigma_0^2$, which takes values $0.057$, $0.33$, and $1.18$ respectively for increasing $M_{\rm max}$. Only the $M_{\rm max} = 10^8\,\mathrm{M}_\odot$ case reaches order unity. The analytical transfer function $\chi_*$, however, effectively assumes $\sigma \approx \sigma_0$ throughout the stream's history. Since $\chi_*$ depends on $\sigma$ through an exponent $\sim \exp[-k^2 \sigma^2 t^2]$, the mismatch between the true and assumed dispersion is amplified at small scales, in this case at the $k \gtrsim 5$~kpc$^{-1}$ region where the discrepancy appears. 

\subsubsection{Realistic simulations} \label{subsubsec:realistic}

Having validated the velocity injection implementation in the idealized setting, we now turn to full realistic simulations that include orbital dynamics and stream formation. 
We adopt a similar physical setup as in Ref.~\cite{Bovy_2016} and Ref.~\cite{Delos_2022} for a GD-1-like stream. The host potential is a flattened logarithmic potential with circular velocity $v_c = 220$~km\,s$^{-1}$ and axial flattening $q_z = 0.9$. The stream progenitor's present-day phase-space coordinates are $(x, y, z) = (12.4, 1.5, 7.1)$~kpc and $(v_x, v_y, v_z) = (107, -243, -105)$~km\,s$^{-1}$, chosen to reproduce a GD-1-like stream at the present time $t=0 \rm{\ Myr}$ \cite{Bovy_2014}. The progenitor velocity dispersion is set to $\sigma_p = 0.365\,(4.5\,\mathrm{Gyr}/t_{\rm age})$~km\,s$^{-1}$, where the $1/t_{\rm age}$ scaling ensures that the stream reaches approximately the same physical length regardless of age, allowing $t_{\rm age}$ to be varied while keeping the length of the stream, and hence the spatial scales considered, fixed to first order. The substructure environment is the same as in the idealized simulations.  
For illustration purposes, the orbital diagnostics and stream post-processing presented in this section use $N_* = 26000$ stars (Figs.~\ref{fig:orbital_quantities} and ~\ref{fig:stream}), consistent with the estimated initial star count for GD-1 \cite{de_Boer_2020}. For the comparison with analytical predictions  (Fig.~\ref{fig:comparison}), we use $N_* = 10^6$ stars to suppress the shot noise floor and enable a clean comparison across a wide range of scales.

\begin{figure}[t]
    \centering
    \includegraphics[width=0.85\linewidth]{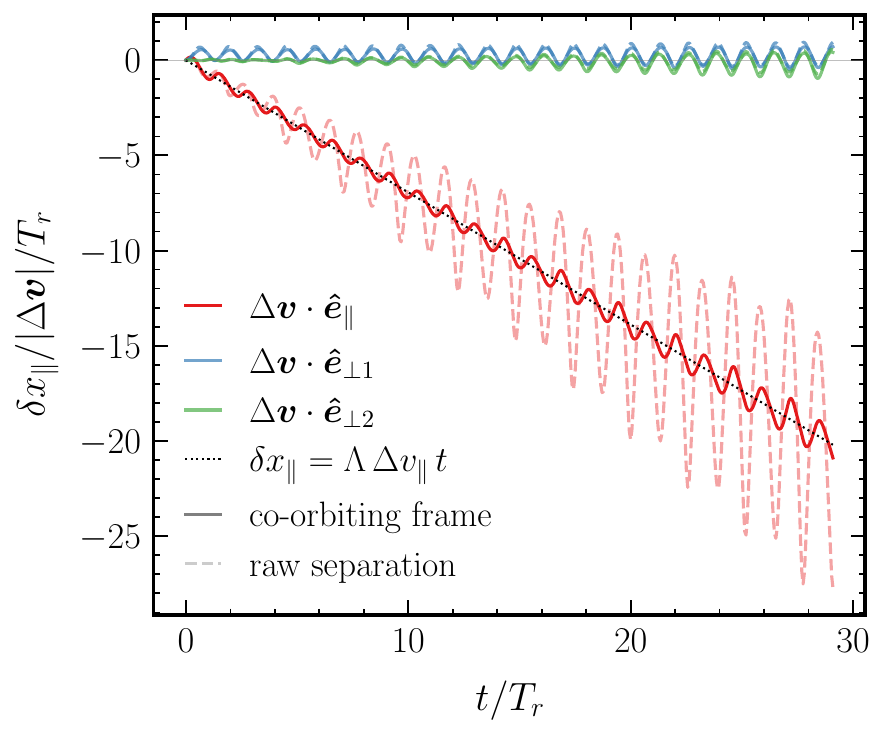}
    \caption{Evolution of the position perturbation $\delta x_\parallel / |\Delta \boldsymbol{v}|$ along the orbital velocity, normalized by the radial orbit period $T_r$, in response to a single velocity kick $\Delta\boldsymbol{v}$ applied at $t = 0 \rm{\ Myr}$ to a star on the unperturbed progenitor orbit. Three kick directions are shown: parallel to the orbital velocity ($\hat{\boldsymbol{e}}_\parallel$, red) and along two perpendicular directions ($\hat{\boldsymbol{e}}_{\perp 1}$, blue; $\hat{\boldsymbol{e}}_{\perp 2}$, green). Only the parallel kick induces a secular drift. Dashed lines show the raw separation between perturbed and unperturbed orbits, whose oscillation amplitude grows due to the periodic expansion and contraction of co-orbiting material. Solid lines show the separation in the co-orbiting frame, rescaled by $\langle v_{\rm orb}\rangle / v_{\rm orb}$, showing a clean secular trend matching Eq.~\eqref{eq:lambda} (dotted).}
    \label{fig:secular}
\end{figure}

We first verify that our simulator correctly reproduces the expected behaviour of orbit perturbations. Fig.~\ref{fig:secular} shows the time evolution of the position perturbation $\delta x_\parallel / |\Delta \boldsymbol{v}|$ along the orbital velocity, normalized by the radial orbit period $T_r$, in response to a single 3D velocity kick $\Delta\boldsymbol{v}$ applied to a star on the unperturbed progenitor orbit. As expected, only the velocity kick component parallel to the orbital velocity ($\Delta \boldsymbol{v}\cdot \hat{\boldsymbol{e}}_\parallel$) induces a secular drift in $\delta x_\parallel$, while the perpendicular kicks produce oscillatory (epicyclic) perturbations that remain bounded over many orbital periods.
The raw separation between perturbed and unperturbed orbits (dashed lines) exhibits growing oscillation amplitudes due to the periodic expansion and contraction of co-orbiting material in proportion to the orbital velocity. Once this effect is factored out by rescaling separations by $\langle v_{\rm orb} \rangle / v_{\rm orb}$ (solid lines), the secular trend becomes clean and matches the linear relation from Eq.~\eqref{eq:lambda}. 

\begin{figure*}[t]
    \centering
    \includegraphics[width=\linewidth]{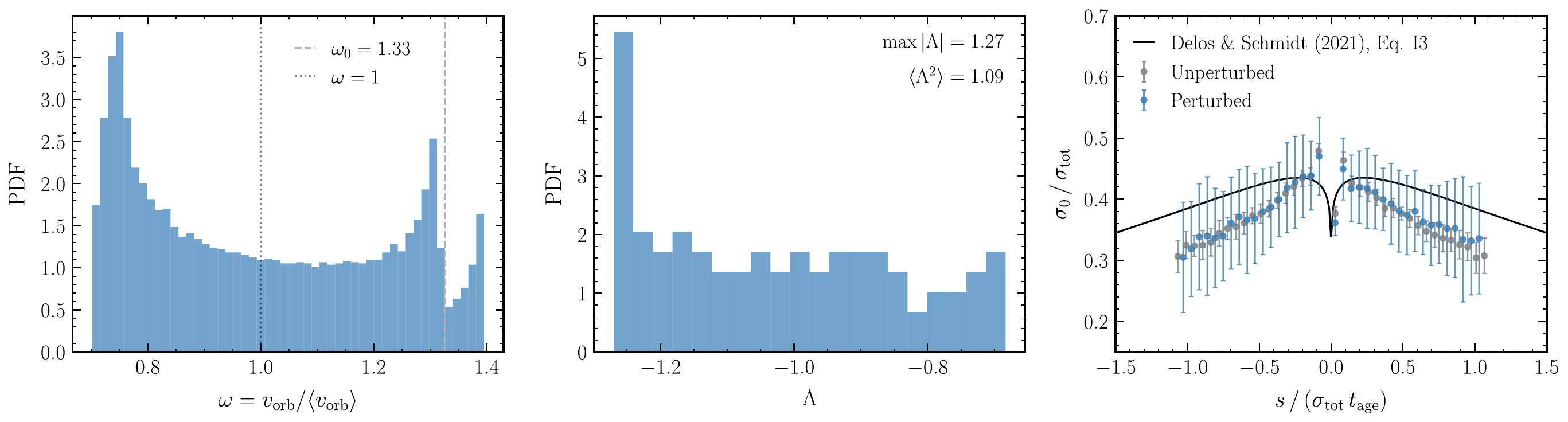}
    \caption{Quantities used as inputs to the analytical predictions of Sec.~\ref{subsec:orbit}. \textit{Left:} distribution of the co-scaling factor $\omega \equiv v_{\rm orb}/\langle v_{\rm orb}\rangle$ sampled over the stream age. The dashed vertical line marks the present-day value $\omega_0 = 1.33$, and the dotted line indicates $\omega = 1$. \textit{Center:} distribution of the kick efficiency $\Lambda$ (see Eq.~\eqref{eq:lambda}), obtained by applying velocity kicks at uniformly sampled orbital phases and measuring the resulting secular drift rate. \textit{Right:} ratio of the local to global velocity dispersion $\sigma_0/\sigma_{\rm tot}$ as a function of normalized position $x/(\sigma_{\rm tot}\,t_{\rm age})$ along the stream. The black curve shows the analytical prediction of Ref.~\cite[Eq.~I3]{Delos_2022}. Grey and blue markers show unperturbed and perturbed ($M_{\rm max} = 10^{7.5}\,\mathrm{M}_\odot$) simulations respectively, averaged over 25 realizations, with error bars indicating the standard deviation across realizations. Both arms of the stream are shown.}
    \label{fig:orbital_quantities}
\end{figure*}

To correctly compare simulations to analytical predictions, we must apply the analytical predictions corrected by the effects of orbital dynamics and stream formation (Sec.~\ref{subsec:orbit}). First, we must extract the kick efficiency $\Lambda$ and the co-scaling factor $\omega \equiv v_{\rm orb}/\langle v_{\rm orb} \rangle$ from the progenitor orbit. Fig.~\ref{fig:orbital_quantities} shows the distributions of both quantities sampled over the stream age. The co-scaling factor $\omega$ (left panel) ranges from $0.70$ to $1.38$, with a present-day value of $\omega_0 = 1.33$ and a time-averaged orbital velocity of $\langle v_{\rm orb} \rangle \simeq 215$~km\,s$^{-1}$. When evaluating the velocity-injection power spectrum via Eq.~\eqref{eq:omega}, we use the full sampled distribution rather than its summary statistics.
The kick efficiency $\Lambda$ (center panel) is obtained by applying velocity kicks at uniformly sampled orbital phases and measuring the resulting secular drift rate (see dotted line in Fig.~\ref{fig:secular}); its distribution yields $\max|\Lambda| = 1.27$ and $\langle \Lambda^2 \rangle = 1.09$, the latter entering as a multiplicative factor in the velocity-injection power spectrum. 

The final ingredient needed for the analytical predictions is the local velocity dispersion $\sigma_0$, which enters the transfer functions $\chi_*$ in Eq.~\eqref{eq:pdensity}. As derived in Ref.~\cite[App.~I]{Delos_2022}, $\sigma_0$ is smaller than the global velocity dispersion $\sigma_{\rm tot}$ of the stream due to self-sorting: stars with similar velocities tend to arrive at similar positions along the stream, narrowing the local velocity spread. Under the assumption that the total velocity distribution is Gaussian, Ref.~\cite{Delos_2022} obtained the ratio $\sigma_0/\sigma_{\rm tot}$ as a function of position along the stream, finding $\sigma_0 \simeq 0.4\,\sigma_{\rm tot}$ for well-populated regions. The right panel of Fig.~\ref{fig:orbital_quantities} compares our simulations, averaged over 25 realizations with $M_{\rm max} = 10^{7.5}\,\mathrm{M}_\odot$, to this prediction. For each realization, we measure the drift velocity $v_\parallel = (s - s_{\rm prog})/(t_{\rm obs} - t_{\rm strip})$ of each star, where $s$ is the one-dimensional arc-length coordinate along the stream track and $s_{\rm prog}$ is the progenitor position (see e.g.~the right panel of Fig.~\ref{fig:stream}). We then compute the global dispersion $\sigma_{\rm tot}$ across both arms and the local dispersion $\sigma_0$ in bins of arc-length, normalizing the bin positions and the local dispersion by $\sigma_{\rm tot}\,t_{\rm age}$ and $\sigma_{\rm tot}$ respectively. Both the unperturbed (grey) and perturbed (blue) simulations are consistent with the analytical curve, with the perturbed simulations showing more scatter due to the perturbations. We notice (based on additionally performed tests) that the scatter increases with $M_{\rm max}$, as expected from the growing contribution of stronger kicks. When comparing simulations to analytical predictions, we compute the mean local dispersion $\sigma_0$ for each realization and then average across realizations; this averaged $\sigma_0$ is used as input to the analytical formulae.

\begin{figure*}[t]
    \centering
    \includegraphics[width=\linewidth]{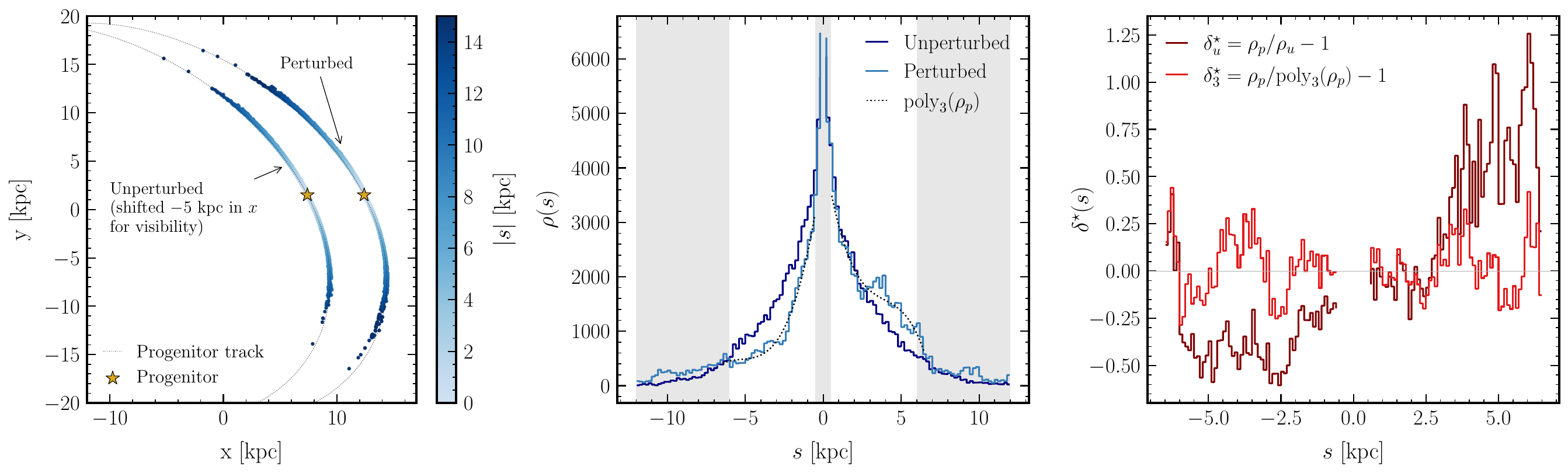}
    \caption{Post-processing of a single stream simulation into a linear density contrast field, for a realization with $t_{\rm age} = 9$~Gyr, $\Delta t=180$~Myr, and $M_{\rm max} = 10^{7.5}\,\mathrm{M}_\odot$. \textit{Left:} perturbed and unperturbed streams in the $x$--$y$ plane of the Galactic frame, colored by arc-length $|s|$ from the progenitor (star symbol). The unperturbed stream is shifted by $-5$~kpc in $x$ for visibility. The dotted grey curve shows the progenitor orbital track. \textit{Center:} linear density $\rho(s)$ as a function of arc-length for the perturbed (light blue) and unperturbed (dark blue) streams. The dotted curve shows the third-order polynomial fit to the perturbed density. Shaded regions indicate excluded zones: the low-density tails and a $\pm 0.5$~kpc buffer around the progenitor. \textit{Right:} two definitions of the density contrast computed from the perturbed stream. The unperturbed-based estimator $\delta^*_u = \rho_{\rm p}/\rho_{\rm u} - 1$ (dark red) uses the matched unperturbed stream to remove the large-scale density profile. The polynomial-based estimator $\delta^*_3 = \rho_{\rm p}/\mathrm{poly}_3(\rho_{\rm p}) - 1$ (red) fits and divides out a smooth baseline from the perturbed density alone, as would be done in observational analyses.}
    \label{fig:stream}
\end{figure*}

With the inputs to the analytical predictions in place, we now turn to the simulation outputs. Fig.~\ref{fig:stream} illustrates a single stream simulation and its post-processing into a linear density contrast field, for a realization with $t_{\rm age} = 9$~Gyr, $\Delta t=180$~Myr and $M_{\rm max} = 10^{7.5}\,\mathrm{M}_\odot$. For a given realization, we simulate two streams with matched initial conditions: one perturbed by the substructure environment and one unperturbed. Star positions are projected onto the one-dimensional arc-length coordinate $s$ along the stream track, that we define, for simplicity, by integrating the progenitor orbit. The left panel shows the perturbed and unperturbed streams in the $x$--$y$ plane of the  Galactic frame, colored by arc-length from the progenitor. The stream arms span on average $\sim 14$~kpc in total, but the highest-density region is concentrated within $\sim 6.5$~kpc of the progenitor; we focus on this region and additionally exclude a $\pm 0.5$~kpc buffer around the progenitor to avoid recently stripped stars. The central panel shows the linear density $\rho(s)$ of both streams, with shaded regions marking the excluded zones. To compute the density contrast $\delta^*$ (right panel), we consider two approaches. The first defines $\delta^*_u(s) = \rho_{\rm p}(s)/\rho_{\rm u}(s) - 1$, using the initial-conditions-matched unperturbed stream to directly remove the large-scale density profile \cite{Bovy_2016}. The second approach to computing the density contrast, applicable also to observational data where a fictitious unperturbed stream is not accessible, fits a third-order polynomial to the perturbed density and defines $\delta^*_3(s) = \rho_{\rm p}(s)/\mathrm{poly}_3(\rho_{\rm p}) - 1$ \cite[e.g.][]{Bovy_2016, Banik_2021a}. As visible in the right panel of Fig.~\ref{fig:stream}, the two estimators produce different density contrast profiles, reflecting the different baseline removal approaches, but both capture the same underlying perturbations. To obtain the power spectrum from a given density contrast field, the field is multiplied by a Hann window function to suppress spectral leakage from the finite stream length before computing the power spectrum. 

\begin{figure*}[t]
    \centering
    \includegraphics[width=\linewidth]{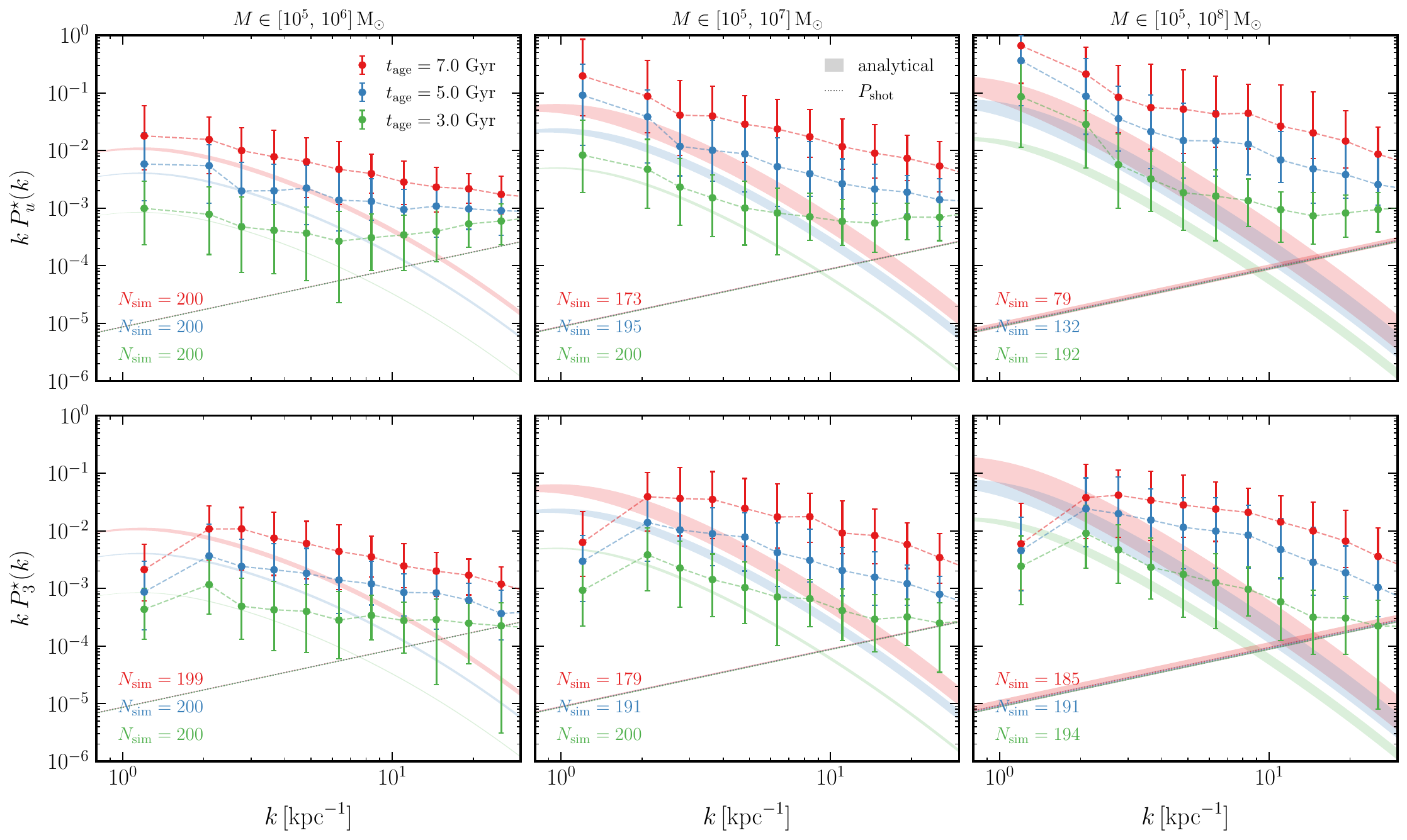}
    \caption{Comparison of the shot noise-subtracted stream density power spectrum $kP^*(k)$ from realistic simulations against analytical predictions, for both density contrast estimators (rows), three substructure mass ranges (columns), and three stream ages (different colors). \textit{Upper row:} unperturbed-based estimator $\delta^*_u$. \textit{Lower row:} polynomial-based estimator $\delta^*_3$. Points show the median power across $N_{\rm sim}$ retained realizations at 11 logarithmically spaced wavenumber bins, with error bars spanning the 16th--84th percentiles. Shaded bands bracket the analytical predictions computed with $D = D(2\pi/L)$ using $\sigma_0 - \sigma_{0,\rm std}$ and $D = 0$ using $\sigma_0 + \sigma_{0,\rm std}$, where $\sigma_{0,\rm std}$ is the standard deviation of the local velocity dispersion across realizations. Dotted lines indicate the median shot noise floor $P_{\rm shot} = L/N_{\star, \rm sel}$ across realizations, where $N_{\star,\rm sel}$ is the number of stars falling within the selected stream region. The number of retained simulations $N_{\rm sim}$ (out of 200) is reported in each panel for each stream age; realizations where $|\delta^*| > 5$ are discarded for this comparison as the stream is too perturbed to enable a sensible comparison with the analytical prescription.}
    \label{fig:comparison}
\end{figure*}

Fig.~\ref{fig:comparison} shows the comparison between the full realistic simulations and the analytical predictions of Sec.~\ref{sec:diffusion}, for both density contrast estimators, three substructure mass ranges, and three stream ages. The upper row shows results for the unperturbed-based estimator $\delta^*_u$, and the lower row for the polynomial-based estimator $\delta^*_3$. Columns correspond to mass ranges $M \in [10^5, 10^6]$, $[10^5, 10^7]$, and $[10^5, 10^8]\,\mathrm{M}_\odot$ from left to right, and colors indicate stream ages $t_{\rm age} = 7$, $5$, and $3$~Gyr. Each data point shows the median power across simulations at 11 logarithmically spaced wavenumber bins, with error bars spanning the 16th--84th percentiles.

The analytical bands bracket the predictions computed with $D = D(2\pi/L)$ using $\sigma_0 - \sigma_{0,\rm std}$ and $D = 0$ using $\sigma_0 + \sigma_{0,\rm std}$, where $\sigma_{0,\rm std}$ is the standard deviation of $\sigma_0$ across realizations. This is intended to capture the combined uncertainty in both the diffusion coefficient and the local velocity dispersion within the analytical framework, given the approximations discussed in Sec.~\ref{subsec:orbit}.

Each panel also reports the number of simulations $N_{\rm sim}$ retained for each age, out of 200 total. Realizations where the maximum density contrast exceeds $|\delta^*| > 5$ are discarded as heavily disrupted streams where the analytical prediction is no more applicable. The number of retained simulations decreases with increasing mass range and stream age, as stronger or more prolonged perturbations leave a larger fraction of realizations 
heavily disrupted.

The colored dotted lines indicate the shot noise floor for each age and mass range. For the polynomial-based estimator, this is the irreducible Poisson floor $P_{\rm shot} = L/N_{\star}^{\rm sel}$, where $N_{\star}^{\rm sel}$ is the number of stars falling within the selected arm region of 6 kpc. For the unperturbed-based estimator, since both streams share the same initial star positions and velocities, the density fluctuations in numerator and denominator are correlated and the noise is no longer purely Poissonian. In all panels, the dotted lines represent the median of $P_{\rm shot}$ across realizations, as $N_{\star}^{\rm sel}$ varies slightly between realizations depending on how many stars end up within the selected arm region.

The polynomial-based estimator shows a pronounced drop in power at the lowest wavenumber bin in each panel. This is expected: the third-order polynomial fit absorbs large-scale density variations, suppressing power at scales comparable to or larger than the polynomial's characteristic scale. 

Both estimators are consistent with the analytical predictions within the error bars at large scales. At intermediate and small scales, both estimators consistently show excess power above the analytical bracket. The analytical predictions of Ref.~\cite{Delos_2022} were previously validated against simulations using the \texttt{streampepperdf} framework of Ref.~\cite{Bovy_2016}, finding good agreement (see Ref.~\cite[Fig.~6]{Delos_2022}); the systematic excess we observe here may therefore reflect additional effects captured by our full 6D treatment that is absent from the one-dimensional simulations used in that validation. The two simulation approaches differ fundamentally in how stream dynamics are modelled. Ref.~\cite{Bovy_2016} works entirely in one-dimensional action-angle space: the stream is described as a PDF $p(\Delta\Omega_\parallel, \Delta\theta_\parallel)$, perpendicular phase-space components are neglected by construction, and velocity kicks are computed only along the mean track. Our simulations instead are more realistic, integrating the full six-dimensional phase-space trajectories of stream stars under velocity kicks sourced by the three-dimensional density field, and naturally retaining perpendicular dispersion and the full orbital response. Several physical effects present in our treatment are suppressed in the one-dimensional framework: (i) cross-terms between parallel and perpendicular kick components, which are small but non-zero; (ii) no linearized coordinate transformation $(\Omega,\theta)\to(\boldsymbol{x}, \boldsymbol{v})$, necessary when converting model predictions to observable coordinates in the phase-space formulation, is required; and (iii) the finite transverse width of the stream, which contributes a small smearing when measuring the 1D density.
Establishing to what extent any of these effects accounts for the observed excess in power at small scales would require additional dedicated tests, which we leave to future work.

\section{Forecasts} \label{sec:forecasts}

The forward model developed in Sec.~\ref{sec:simulations} provides a fast, differentiable mapping from substructure parameters to observable stream statistics. Having validated it against analytical predictions, we now exploit its differentiability to forecast the sensitivity of a GD-1-like stream to substructure power spectrum parameters in different scenarios.
A key result demonstrated here for the first time is the expected gain in constraining power on substructure by using velocity statistics jointly with stellar density.

\subsection{Setup} \label{subsec:forecast_setup}

We adopt the same GD-1-like stream setup as in Sec.~\ref{subsubsec:realistic}, with $t_{\rm age} = 5$~Gyr and $N_\star = 1700$ stars, consistent with the spectroscopic sample of GD-1 from Ref.~\cite{Tavangar_2025}. We use $\Delta t = 100$~Myr, $\eta = 0.05$~kpc, and $N_\delta = 1$.

The substructure environment is described by the power spectrum of Eq.~\eqref{eq:Pksub}, with subhalo density profiles modelled as NFW spheres truncated at the virial radius, whose Fourier transform is taken from Ref.~\cite{Cooray_2002}. The scale radius is set by $R_s = [M/(4\pi\rho_s f(c))]^{1/3}$, where $f(c) = \ln(1+c) - c/(1+c)$, with scale density $\rho_s = 5\times 10^7\,\mathrm{M}_\odot\,\mathrm{kpc}^{-3}$ and concentration $c = 20$. The scale radius is set by the mass--concentration relation $R_s = [M/(4\pi\rho_s f(c))]^{1/3}$, where $f(c) = \ln(1+c) - c/(1+c)$. 

\begin{figure*}[t]
    \centering
    \includegraphics[width=\linewidth]{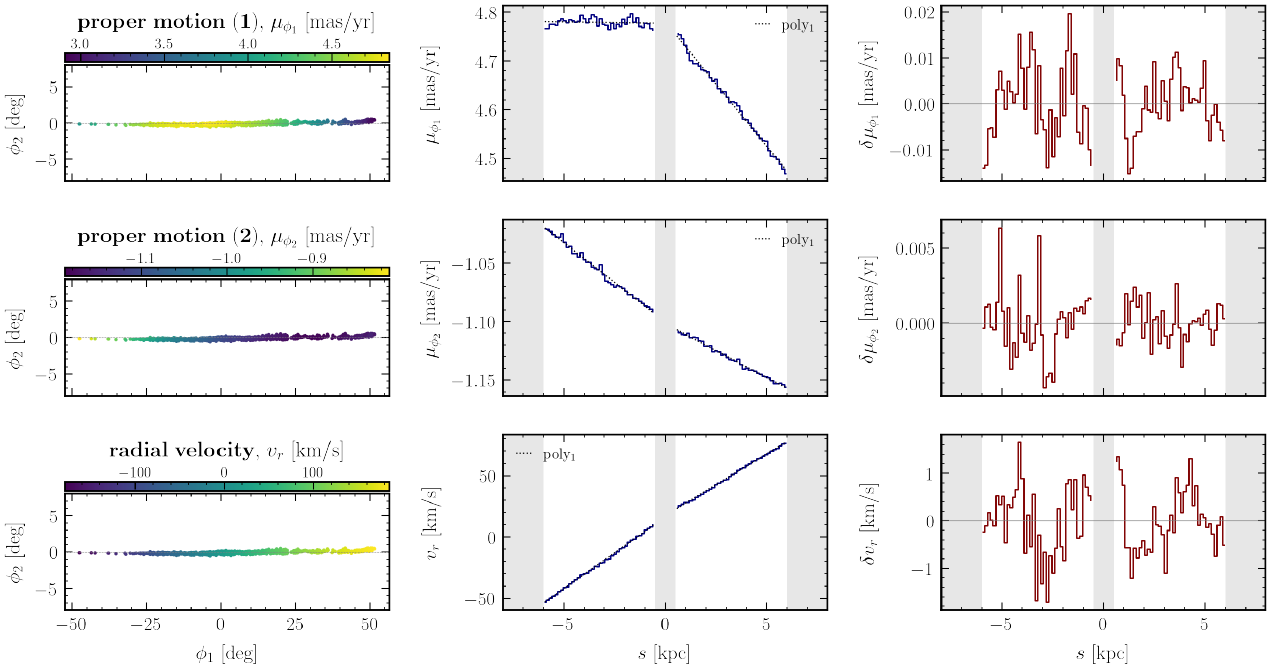}
    \caption{Construction of the kinematic contrast fields for a single stream realization ($t_{\rm age} = 5$~Gyr, $M_{\rm hm} = 10^6\,\mathrm{M}_\odot$). Rows show, from top to bottom, the two proper motion components, $\mu_{\phi_1}$ and $\mu_{\phi_2}$, and the radial velocity, $v_r$. \textit{Left:} stars in the on-sky stream frame $(\phi_1, \phi_2)$, colored by the corresponding kinematic quantity. \textit{Center:} the one-dimensional profile of each field binned along arc-length $s$ (blue), with the first-order polynomial baseline overlaid. Shaded regions mark the excluded low-density tails and the $\pm0.5$~kpc buffer around the progenitor. \textit{Right:} the resulting contrast fields, $\delta = \mathrm{field} - \mathrm{poly}_1(\mathrm{field})$ for the kinematic fields, from which the power spectra are computed.
    }
    \label{fig:observables}
\end{figure*}

The mean substructure density is modelled as $\bar{\rho} = f_{\rm sub}\,\bar{\rho}_{\rm MW}$, evaluated at the mean Galactocentric radius of the GD-1-like orbit, $r \simeq 20$~kpc. Modelling the Milky Way dark matter halo as an NFW profile with $M_{200} = 1.3\times10^{12}\,\mathrm{M}_\odot$ \cite{McMillan_2016} gives $\bar{\rho}_{\rm MW}(20\,\mathrm{kpc}) \approx 2\times10^6\,\mathrm{M}_\odot\,\mathrm{kpc}^{-3}$. The local subhalo mass fraction $f_{\rm sub}(r)$ is taken from the Aquarius simulations \cite{Springel_2008_Aquarius}, which gives $f_{\rm sub}(20\,\mathrm{kpc}) \approx 0.004$. We hence use a fiducial mean substructure density of $\bar{\rho} \approx 8\times10^3\,\mathrm{M}_\odot\,\mathrm{kpc}^{-3}$. The radial variation of $\bar{\rho}$ along the orbit is neglected for simplicity, as the GD-1-like orbit considered here spans $r \in [13.5, 26.2]$~kpc. However, a radial dependence of the substructure power spectrum is straightforward to include in the model.

We consider a subhalo mass function of the form 
\begin{equation} \label{eq:mf_forecast} 
\frac{\mathrm{d}n}{\mathrm{d}M}(M) = A\, M^{-\alpha} \left(1 + \gamma\,\frac{M_{\rm hm}}{M}\right)^{-\beta}, 
\end{equation} 
where $\alpha$ is the CDM power-law slope and the second factor introduces a suppression below the half-mode mass $M_{\rm hm}$, following Ref.~\cite{Schneider_2012}. Dark matter models with non-negligible thermal velocities, such as warm dark matter (WDM) \cite{Hogan_2000_WDM}, predict a cutoff in the matter power spectrum that suppresses the formation of low-mass halos below a characteristic scale. The half-mode mass, $M_{\rm hm}$, parametrizes this cutoff: it corresponds to the mass enclosed within the scale at which the WDM transfer function falls to half the CDM value, and maps one-to-one onto the WDM particle mass (within a given freeze-out scenario). Heavier particles correspond to smaller $M_{\rm hm}$ and hence less suppression, recovering canonical CDM in the limit $M_{\rm hm} \to 0$. The shape parameters $\beta = 1.16$ and $\gamma = 2.7$ are fixed following Ref.~\cite{Schneider_2012}. The normalization $A$ is determined by requiring the integrated subhalo mass density to equal the prescribed mean substructure density $\bar{\rho}$. The mass integration spans $M_{\rm min} = 10^4\,\mathrm{M}_\odot$ to $M_{\rm max} = 10^8\,\mathrm{M}_\odot$.

The substructure velocity distribution is an isotropic Maxwellian with scale velocity $u_0 = 120$~km\,s$^{-1}$.

\subsection{Observables} \label{subsec:forecast_observables}

Most previous works constraining dark matter substructure from stellar streams have relied solely on the stream density power spectrum \cite[e.g.][]{Banik_2021a, Banik_2021b}. Only recently has kinematic information been incorporated, with Ref.~\cite{Nibauer_2026} deriving the first constraints from the radial velocity dispersion of GD-1 using 160 member stars. The observational landscape is rapidly evolving: Gaia provides precise proper motions for GD-1 members across the full stream \cite{Tavangar_2025}, and a recent DESI DR2 catalog presents 608 spectroscopically confirmed GD-1 members with homogeneous radial velocity measurements \cite{Jarvis_2026}. Looking ahead, dedicated facilities such as the VIA Project \cite{VIA_project}, which is building spectrographs targeting stellar stream kinematics, will provide richer kinematic information for a larger number of stream members.

Motivated by this, we consider the full set of kinematic observables accessible along the stream. The simulated 3D galactocentric positions and velocities are first converted to the heliocentric frame using the Sun's galactocentric position $\boldsymbol{r}_\odot = (-8.122, 0, 0.021)$~kpc and velocity $\boldsymbol{v}_\odot = (12.9, 245.6, 7.78)$~km\,s$^{-1}$, following the \texttt{astropy} \cite{astropy} galactocentric defaults. Each star is then projected onto a local stream frame, defined by a right-handed orthonormal triad $(\hat{\boldsymbol{e}}_{\phi_1}, \hat{\boldsymbol{e}}_{\phi_2}, \hat{\boldsymbol{n}})$ computed at each point along the progenitor orbit, where $\hat{\boldsymbol{e}}_{\phi_1}$ is the along-stream tangent on the heliocentric unit sphere, $\hat{\boldsymbol{e}}_{\phi_2}$ the cross-stream direction, and $\hat{\boldsymbol{n}}$ the line-of-sight. This yields three kinematic observables per star,
\begin{equation}
    \mu_{\phi_1} = \frac{\boldsymbol{v}_h \cdot 
    \hat{\boldsymbol{e}}_{\phi_1}}{r}, \quad
    \mu_{\phi_2} = \frac{\boldsymbol{v}_h \cdot 
    \hat{\boldsymbol{e}}_{\phi_2}}{r}, \quad
    v_r = \boldsymbol{v}_h \cdot \hat{\boldsymbol{n}},
\end{equation}
where $\boldsymbol{v}_h$ is the heliocentric velocity and $r$ the 
heliocentric distance.

Each kinematic field is binned along the arc-length coordinate $s$ to obtain a one-dimensional profile, using the same 6~kpc high-density arm region and $\pm0.5$~kpc progenitor exclusion buffer as in Sec.~\ref{subsubsec:realistic}. For the density, we use the polynomial-based estimator $\delta^*_3$. For the kinematic fields, large-scale trends are removed by fitting and subtracting a first-order polynomial baseline, $\delta\mu_{\phi_1} = \mu_{\phi_1} - \mathrm{poly}_1(\mu_{\phi_1})$, and similarly for $\mu_{\phi_2}$ and $v_r$, in analogy with the polynomial-based density estimator $\delta^*_3$. Fig.~\ref{fig:observables}, similarly to Fig.~\ref{fig:stream}, illustrates this procedure for a single stream realization, showing each kinematic observable on the sky, its binned profile along arc-length, and the resulting contrast field.

\begin{figure*}[t]
    \centering
    \includegraphics[width=\linewidth]{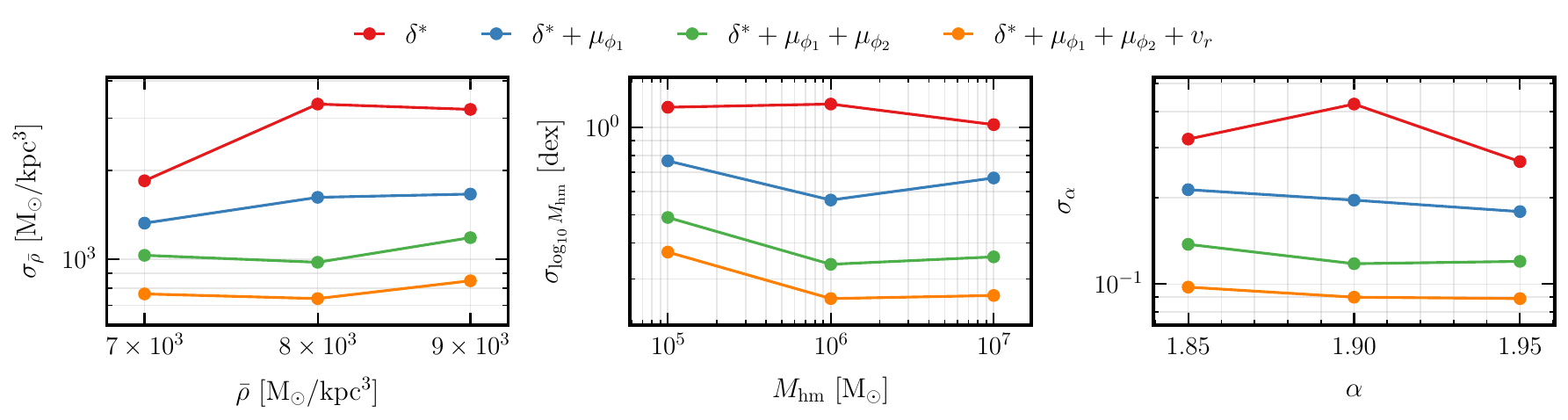}
    \caption{Conditional Fisher constraints on the substructure parameters $\bar{\rho}$ (left), $\log_{10} M_{\rm hm}$ (center), and $\alpha$ (right), as a function of each parameter's fiducial value, with the other two parameters fixed at the center fiducial $(\bar{\rho} = 8\times10^3\,\mathrm{M}_\odot\,\mathrm{kpc}^{-3},\  M_{\rm hm} = 10^6\,\mathrm{M}_\odot,\  \alpha = 1.9)$ and $t_{\rm age} = 5$~Gyr. Each panel shows the forecasted uncertainty $\sigma_\theta$ for four cumulative data vectors of increasing kinematic content: $\delta^*$ alone (red), $\delta^* + \mu_{\phi_1}$ (blue), $\delta^* + \mu_{\phi_1} + \mu_{\phi_2}$ (green), and the full kinematic vector $\delta^* + \mu_{\phi_1} + \mu_{\phi_2} + v_r$ (orange). Each data vector includes all auto- and cross-spectra among its components. Smaller values indicate tighter constraints, with $\sigma_\theta/\theta \ll 1$ meaning the parameter is well constrained rather than merely bounded.}
    \label{fig:fisher}
\end{figure*}

Each contrast field is multiplied by a Hann window to suppress spectral leakage before computing the one-dimensional power spectrum via FFT. Cross-power spectra between fields $A$ and $B$ are defined as the real part of $\langle \hat{A}_k \hat{B}_k^* \rangle$, using the same normalization as the auto-spectra, so that $\langle \hat{A}_k \hat{A}_k^* \rangle$ reduces to the auto-spectrum of $A$.

The full data vector comprises the 10 auto- and cross-power spectra $P_{AB}(k)$ for all pairs $(A, B)$ drawn from $\{\delta^*, \mu_{\phi_1}, \mu_{\phi_2}, v_r\}$, namely $P_{\delta^*\delta^*}$, $P_{\mu_{\phi_1}\mu_{\phi_1}}$, $P_{\mu_{\phi_2}\mu_{\phi_2}}$, $P_{v_r v_r}$, $P_{\delta^*\mu_{\phi_1}}$, $P_{\delta^*\mu_{\phi_2}}$, $P_{\delta^* v_r}$, $P_{\mu_{\phi_1}\mu_{\phi_2}}$, $P_{\mu_{\phi_1} v_r}$, and $P_{\mu_{\phi_2} v_r}$.
The cross-spectra carry additional information beyond the auto-spectra, encoding correlations between density and velocity perturbations induced by the same substructure kicks.

For the forecasts presented here, we assume full kinematic coverage: all $N_\star = 1700$ stars are assumed to have measured positions, proper motions $\mu_{\phi_1}$ and $\mu_{\phi_2}$, and radial velocity $v_r$. This is not yet the case for GD-1, but represents the target regime of ongoing and upcoming surveys, and allows us to assess the information gain from each additional kinematic component.

\subsection{Fisher forecast results} \label{subsec:forecast_results}

\begin{figure*}[t]
    \centering
    \includegraphics[width=\linewidth]{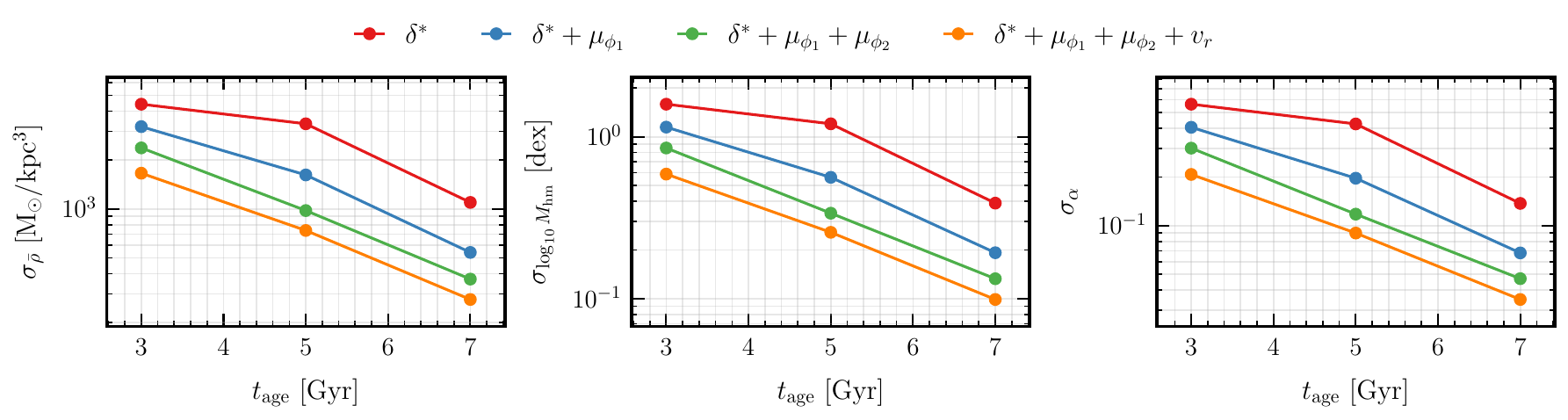}
    \caption{Same as Fig.~\ref{fig:fisher}, but as a function of stream age $t_{\rm age} \in \{3, 5, 7\}$~Gyr at the center fiducial $(\bar{\rho} = 8\times10^3\,\mathrm{M}_\odot\,\mathrm{kpc}^{-3},\  M_{\rm hm} = 10^6\,\mathrm{M}_\odot,\  \alpha = 1.9)$.}
    \label{fig:fisher_tage}
\end{figure*}

For each configuration, we run $N_{\rm sim} = 30000$ realizations and discard those with $|\delta^*| > 5$ as heavily disrupted streams, following the criterion of Sec.~\ref{subsubsec:realistic}. Each kinematic and density field is binned into 20 arc-length bins along $s$, yielding power spectra with 10 wavenumber bins (the $k=0$ bin 
is excluded). Power spectra are computed separately for the leading and trailing arms and concatenated into a single data vector. 
Depending on the chosen observable target (for instance $\delta^*$ alone, or the full joint $\{\delta^*, \mu_{\phi_1}, \mu_{\phi_2}, v_r\}$ together with all its cross-spectra,  as defined in Sec.~\ref{subsec:forecast_observables}), the corresponding spectra are concatenated across arms into a single data vector $\boldsymbol{d}$, and the empirical covariance matrix $\boldsymbol{C}$ is estimated from $\boldsymbol{d}$ across the surviving realizations. This single covariance matrix encodes all bin-to-bin, field-to-field, and arm-to-arm correlations relevant for the chosen target.

We compute the Fisher information matrix as 
\begin{equation} 
    F_{ij} = \frac{\partial \boldsymbol{\mu}^\top}{\partial \theta_i}\, \boldsymbol{C}^{-1}\, \frac{\partial \boldsymbol{\mu}}{\partial \theta_j},    
\end{equation} 
where $\boldsymbol{\mu}(\theta)$ is the mean data vector, $\boldsymbol{C}$ is the empirical covariance estimated from the realizations at the fiducial point, and $\boldsymbol{\theta} = (\bar{\rho}, M_{\rm hm}, \alpha)$. Derivatives are computed via \jax automatic differentiation through the full forward model, including the stream simulator, the projection onto stream-frame observables, and the power spectrum estimators; we have verified that the resulting gradients agree with finite-difference derivatives to numerical precision. We report conditional (unmarginalized) Fisher uncertainties $\sigma_{\theta_i} = 1/\sqrt{F_{ii}}$, which assume the other two parameters are perfectly known. The stream age $t_{\rm age}$ is held fixed in the Fisher analysis.

We perform two sweeps. First, we vary each substructure parameter around the center fiducial $(\bar{\rho} = 8\times10^3\,\mathrm{M}_\odot\, \mathrm{kpc}^{-3},\ M_{\rm hm} = 10^6\,\mathrm{M}_\odot,\ \alpha = 1.9)$ at fixed $t_{\rm age} = 5$~Gyr, recomputing the covariance and derivatives at each point. Second, we vary the stream age $t_{\rm age} \in \{3, 5, 7\}$~Gyr at the center fiducial. We parametrize the cutoff through $\log_{10} M_{\rm hm}$ rather than $M_{\rm hm}$, since the half-mode mass spans several orders of magnitude; this also matches how constraints are reported in the literature. In both sweeps, we evaluate four cumulative data vectors built from progressively richer kinematic information: $\delta^*$ alone (1 auto-spectrum), $\delta^* + \mu_{\phi_1}$ (2 auto + 1 cross), $\delta^* + \mu_{\phi_1} + \mu_{\phi_2}$ (3 auto + 3 cross), and the full $\delta^* + \mu_{\phi_1} + \mu_{\phi_2} + v_r$ (4 auto + 6 cross). 

Fig.~\ref{fig:fisher} shows the Fisher uncertainty $\sigma_\theta$ on each parameter as a function of its fiducial value. Adding kinematic observables monotonically tightens constraints on all three parameters: the full kinematic vector improves constraints by a factor of $\sim 3$--$5$ relative to density alone, with each successive kinematic component contributing a roughly comparable gain. We find from additional tests that the cross-spectra contribute a genuine part of this gain: at fixed observable content, including the cross-spectra rather than keeping the auto-spectra alone tightens the constraints up to a factor of $\sim 1.5$, with the gain growing as more fields are combined. This confirms that the correlations between density and kinematic perturbations, induced by the same substructure kicks, carry real information.
For $\bar{\rho}$ and $\alpha$, the relative uncertainties of $\sim 10^{-1}$ indicate that these parameters are well constrainable at least individually. For the mass function cutoff, the constraint is most directly read as the absolute precision $\sigma_{\log_{10} M_{\rm hm}}$ in dex shown in the center panel: stellar density alone reaches $\sim 1.2$~dex, localizing the cutoff only to within an order of magnitude, while the full kinematic vector reaches $\sim 0.25$--$0.3$~dex across the explored range, pinning $M_{\rm hm}$ down to within a factor of $\sim 2$. These forecasts therefore indicate that the half-mode mass that can be resolved depends on the available kinematic information: density alone supports only a coarse, order-of-magnitude statement, whereas the full kinematic vector could yield a genuine measurement of the cutoff scale.

Fig.~\ref{fig:fisher_tage} shows the constraints as a function of $t_{\rm age}$. All three parameters are constrained substantially better in older streams, improving by a factor of $\sim 4$--$6$ between $t_{\rm age} = 3$ and $7$~Gyr. This reflects the longer integration time over which substructure perturbations accumulate, increasing the signal-to-noise of the power spectrum measurement. The ordering of the four data vectors is preserved across all stream ages, with kinematic information consistently improving constraints over density alone.

It is instructive to place these forecasts in the context of current constraints on the half-mode mass, while bearing in mind that our numbers are forecasted sensitivities under idealized assumptions (full kinematic coverage of all $N_\star$ stars, no observational uncertainties) rather than measurements.\footnote{Literature constraints also adopt differing priors ($\log M_{\rm hm}$ versus $1/m$), confidence conventions (95\% CL versus 20:1 likelihood ratio), and units ($M_\odot$ versus $M_\odot/h$), so the comparison is necessarily approximate.}

At the central fiducial ($M_{\rm hm} = 10^6\,\ \mathrm{M}_\odot$, $t_{\rm age} = 5$~Gyr), our density-only forecast yields $\sigma_{\log_{10} M_{\rm hm}} \approx 1.2$~dex. This is in reasonably good agreement with the measurement of Ref.~\cite{Banik_2021b}, who obtained $M_{\rm hm} = 10^{6.1\pm1.0}\,\mathrm{M}_\odot$ from the density power spectra of two streams, GD-1 and Pal 5, providing a useful consistency check on our forecasting framework. Among other stream-based constraints, the GD-1 velocity dispersion alone yields $M_{\rm hm} < 10^{8.6}\,\mathrm{M}_\odot$ marginalized over subhalo concentration \cite{Nibauer_2026}.

Including the full kinematic information in our forecasts tightens $\sigma_{\log_{10} M_{\rm hm}}$ to  $\approx 0.26$~dex, and for an older stream ($t_{\rm age} = 7$~Gyr) it reaches $\approx 0.1$~dex. That is, a cutoff at $M_{\rm hm} = 10^6 M_\odot$ would be detectable using this single stellar stream under the optimistic conditions assumed.

For context, the tightest current limits come from strong gravitational lensing of quadruply-imaged quasars. The most recent analysis, using JWST observations of 28 systems, constrains the cutoff to $M_{\rm hm} \lesssim 10^{7.2}$--$10^{7.4}\,\mathrm{M}_\odot$ (depending on the assumed subhalo abundance from $N$-body simulations or semi-analytic models) \citep{Gilman_2026}. This tightens earlier flux-ratio constraints of $M_{\rm hm} < 10^{7.8}\,\mathrm{M}_\odot$ \citep{Hsueh_2019, Gilman_2019} and the JWST warm-dust result $M_{\rm hm} < 10^{7.6}\,\mathrm{M}_\odot$ \citep{Keeley:2024brx}.
Earlier joint analyses combining multiple probes reached comparable limits, with lensing, the Lyman-$\alpha$ forest, and Milky Way satellites together giving $M_{\rm hm} < 3\times10^7\,\mathrm{M}_\odot/h$ \cite{Enzi_2021} and a unified analysis of lensing and the satellite population reaching $M_{\rm hm} < 10^{7.0}\,\mathrm{M}_\odot$ \cite{Nadler_2021}.
All these limits are mutually consistent and disfavour a turnover in the halo mass function above $\sim 10^7\,\mathrm{M}_\odot$. 

We stress that this is not a like-for-like comparison: our forecasts quote a $1\sigma$ relative uncertainty on $\log_{10} M_{\rm hm}$ at a fiducial value, whereas the lensing and joint-analysis results are upper limits under different statistical conventions. Nonetheless, the comparison suggests that a single well-measured stream with full kinematic information could constrain $M_{\rm hm}$ at least as well as these probes, providing a local, purely gravitational measurement with systematics independent of those of lensing and satellite-based methods.

\section{Discussion}  \label{sec:discussion}

We have presented a fast, differentiable forward model for stellar streams in the diffusion regime, validated it against analytical predictions, and demonstrated its use for forecasting sensitivity to dark matter substructure. In this section, we discuss the current limitations of the framework and possible improvements (Sec.~\ref{subsec:limit}), compare our approach to previous work (Sec.~\ref{subsec:litcomparison}), and outline future directions (Sec.~\ref{subsec:outlook}).

\subsection{Current limitations and possible improvements} \label{subsec:limit}

\paragraph*{Stream formation.} We adopt a simplified ejection model with uniform stripping times and a fixed progenitor velocity dispersion. In reality, as discussed in Sec.~\ref{subsec:init}, tidal stripping is episodic and concentrated near pericenter passages, and the progenitor's mass and internal structure evolve over time as it loses stars. More physically motivated stripping prescriptions, calibrated against $N$-body simulations of dissolving clusters, could easily be incorporated without modification to the velocity injection machinery. The main practical caveat is that non-uniform stripping times would change the ordering relative to the kick times across realizations, preventing \texttt{jit} compilation without recompilation for each new draw (see Sec.~\ref{subsec:sim_diffusion}), introducing a non-negligible computational overhead.

\paragraph*{Host galaxy potential.} We adopt a flattened logarithmic potential for simplicity and to enable direct comparison with previous work. More realistic Milky Way potentials, incorporating a bulge, disk, and dark matter halo with observationally calibrated parameters \cite[e.g.][]{McMillan_2016}, can be straightforwardly substituted into the orbital integration via \galax without any modification to the velocity injection framework, at a modest increase in the integration cost. Furthermore, since our approach integrates orbits directly rather than relying on action-angle coordinates, it can naturally accommodate time-varying potentials, which are inaccessible to action-angle-based stream models \cite{Bovy_2016}.

\paragraph*{Spatially varying substructure environment.} The substructure power spectrum and mean density are assumed to be constant along the stream orbit. In practice, the abundance and properties of subhalos vary with Galactocentric radius, and the stream samples different environments as it moves between pericenter and apocenter. Allowing $\mathcal{P}(q)$ and $\bar{\rho}$ to vary as a function of orbital phase, which can be easily implemented in the current setup, would capture this effect.

\paragraph*{Substructure velocity distribution.} We assume an isotropic Maxwellian velocity distribution for the substructure in the Galactic frame. The true velocity distribution of subhalos is expected to be anisotropic, with a radial bias that depends on Galactocentric radius, and may include contributions from coherent streams of recently accreted material. More realistic distributions, measured from simulations or parametrized analytically \cite[e.g.][]{Mao_2013, zhang2026setnightfirebuilding}, can be straightforwardly substituted into the framework.

\paragraph*{Temporal correlations in the substructure field.} At each kick time, we sample an independent realization of the density field $\delta(\boldsymbol{q})$, treating successive kicks as temporally uncorrelated. This is a good approximation when the kick interval $\Delta t$ exceeds the substructure crossing time $(qu)^{-1}$, but breaks down for the largest-scale modes. Introducing temporal correlations between successive density fields would improve the accuracy at large scales, although it requires specifying a temporal correlation kernel that is not given by any standard prescription. An alternative route would be to bypass the statistical sampling entirely and use density fields extracted directly from zoom-in cosmological $N$-body simulations of Milky Way systems, which would naturally include temporal correlations, non-Gaussianity, and realistic spatial clustering of substructure, at the cost of relying on the resolution and modelling assumptions of those simulations.

\paragraph*{Observational effects.} Our current predictions are for the intrinsic stream density and kinematic power spectra. A comparison to data requires modeling the observational pipeline that maps the true stream configuration to the observed one. This includes, for example, measurement uncertainties on astrometric and spectroscopic quantities, incomplete detection of stream stars due to selection effects, and contamination from foreground or background stars that are not removed by the selection process. In principle, these effects can be incorporated as additional differentiable layers in the forward model, so that its output directly predicts the raw observables rather than the intrinsic stream properties.

\paragraph*{Baryonic substructure.} The current framework includes only dark matter substructure, but baryonic structures such as giant molecular clouds, the Galactic bar, and spiral arms can contribute perturbations to stellar streams \cite[e.g.][]{Banik_2021a}. This represents a genuine open issue rather than a technical limitation: the substructure power spectrum of baryonic perturbers from high-resolution hydrodynamical simulations is itself uncertain, being limited by resolution, sensitive to subgrid physics, and time-dependent in ways that are not captured by a static $\mathcal{P}(q)$. Within our framework, baryonic contributions could be included by adding their power spectrum to the dark matter component, or by modelling known baryonic perturbers as additional deterministic perturbations applied alongside the stochastic diffusive kicks. For the purposes of dark matter constraints, the key requirement is a prior over physically plausible baryonic substructure contributions, which is to be marginalized over in the inference of dark matter properties. Devising such a prior remains on open challenge at this point.

\subsection{Comparison to previous works} \label{subsec:litcomparison}

Only a few works so far have studied the regime where stellar streams are subjected to many small-scale perturbations from a population of dark substructures \cite{Bovy_2016, Banik_2021a, Banik_2021b, Delos_2022, Nibauer:2024uue, Nibauer_2026}. We briefly compare these with our work below.

\paragraph*{Delos \& Schmidt (2022) \cite{Delos_2022}.} Our forward model builds on this work, adopting their velocity injection formalism to model the cumulative statistical effect of dark matter substructure as velocity kicks on stream stars, rather than simulating individual encounters. Beyond this shared ingredient, our approach extends their work by coupling the velocity injection formalism to direct numerical orbit integration with full stream formation, avoiding the approximate analytical treatment of orbital dynamics and doing away with the simplifying assumptions discussed in Sec.~\ref{subsec:orbit}.

\paragraph*{Bovy et al. (2017) \cite{Bovy_2016} and Banik et al. (2021a, 2021b) \cite{Banik_2021a, Banik_2021b}.} Ref.~\cite{Bovy_2016} introduced a linear perturbation framework in action-angle space to compute the effect of subhalo flybys on cold tidal streams. Working in one-dimensional action-angle space, the stream is described as a PDF, $p(\Delta\Omega_\parallel, \Delta\theta_\parallel)$, and the perturbed distribution function of each stream segment is reconstructed by undoing the effect of all impacts along the mean track, treating each encounter as impulsive. Building on this, Ref.~\cite{Banik_2021a, Banik_2021b} applied the framework to GD-1 and Pal 5, accounting for baryonic perturbers alongside dark subhalos, and placed constraints on warm and fuzzy dark matter models.
In contrast, our approach models the collective statistical effect of the entire subhalo population through the substructure power spectrum without resolving individual flybys, making it naturally suited to the low-mass regime where the number of encounters becomes large. Additionally, by working in full 6D phase space rather than action-angle coordinates, our simulations are not restricted to static axisymmetric potentials and can straightforwardly accommodate time-varying or non-axisymmetric host potentials, which are inaccessible to action-angle-based frameworks \cite{Bovy_2016}.

\paragraph*{Nibauer et al. (2025a, 2025b) \cite{Nibauer:2024uue, Nibauer_2026}.} Ref.~\cite{Nibauer:2024uue} introduced a framework combining Hamiltonian perturbation theory with differentiable simulations to model the effect of dark matter subhalos on stream observables. By applying perturbation theory on top of a time-evolving background potential, the method works in observable coordinates and supports flexible, non-axisymmetric host potentials, going beyond the static axisymmetric restriction of action-angle approaches. Like our work, it is implemented in \jax and natively supports automatic differentiation.
Building on this framework, Ref.~\cite{Nibauer_2026} derived the first kinematic constraints on the dark matter subhalo population from the radial velocity dispersion of GD-1, constraining subhalo properties in the mass range $10^5$--$10^9\,\mathrm{M}_\odot$. The key distinction from our approach is that individual subhalo encounters are treated discretely, each entering as a separate perturbation computed to linear order and summed \citep{Nibauer:2024uue}. The number of interactions therefore grows with the 
number of perturbers, becoming computationally expensive as $M_{\rm min}$ decreases and the number of encounters grows steeply (see e.g.~Fig.~\ref{fig:runtime} for a rough comparison). Our diffusion-regime framework instead models the collective statistical effect of arbitrarily many low-mass subhalos through the substructure power spectrum, without resolving individual encounters.

\subsection{Outlook} \label{subsec:outlook}

The framework developed in this work opens several roads for future research.

\paragraph*{Alternative summary statistics.} Our forecasts are based on the one-dimensional power spectra of the stream density and kinematic fields, but the differentiable forward model can make predictions for any summary statistic that is a function of the simulated star positions and velocities. Higher-order statistics such as the bispectrum, wavelet coefficients, or learned summary statistics via neural compression could extract additional information.

\paragraph*{Hybrid modeling of massive perturbers.} Our diffusion-regime approach is optimally suited for the cumulative effect of many low-mass subhalos, but rare encounters with massive perturbers ($M \gtrsim 10^8\,\mathrm{M}_\odot$) can produce localized features such as gaps and track deviations that lie outside the diffusion regime. A natural extension is a hybrid scheme that models the low-mass population statistically through the substructure power spectrum, as done here, while resolving individual encounters with the most massive perturbers explicitly. This would combine the computational efficiency of the diffusion-regime treatment at low masses with the accuracy of direct encounter modeling at high masses.

\paragraph*{Connection to gravitational lensing.}
It is worth noting that the power spectrum of the gravitational potential gradient is the quantity that controls both stream perturbations and strong lensing perturbations by dark matter substructure. While the two probes sample different environments and redshifts, they are sensitive to the same underlying substructure statistics. A joint analysis, or at least a consistent parametrization of the substructure power spectrum across both probes, could therefore provide complementary constraints on the nature of dark matter. 

\section{Conclusion}  \label{sec:conclusion}
Stellar streams are sensitive probes of dark matter substructure, recording the cumulative gravitational perturbations from the low-mass halos that populate the Galactic halo. We have presented a fast, differentiable forward model for perturbed stellar streams in the diffusion regime, combining the velocity injection formalism of Ref.~\cite{Delos_2022} with direct numerical orbit integration and an explicit treatment of stream formation. Rather than resolving individual substructure encounters, the simulations imprints the collective statistical effect of the entire population, depending on the environment only through its power spectrum $\mathcal{P}(q)$. We validated the implementation against the analytical predictions in both an idealized setup and a realistic GD-1-like configuration with orbital dynamics and stream formation (Sec.~\ref{sec:simulations}). 

As a first application, we forecast how well a GD-1-like stream can 
constrain the substructure power spectrum, going beyond the stream density to include the full set of kinematic observables: the two proper motions and the radial velocity (Sec.~\ref{sec:forecasts}). We find that kinematic information tightens the forecast constraints by a factor of $\sim 3$--$5$ relative to density alone. For the half-mode mass $M_{\rm hm}$ with fiducial value of $10^6 M_\odot$, this improves the precision from $\sim 1.2$~dex with density alone to $\sim 0.25$~dex with the full kinematic vector at $t_{\rm age} = 5$~Gyr, reaching $\sim 0.1$~dex for an older 7~Gyr stream. These are idealized forecasts, but they suggest that a single well-measured stream could constrain the free-streaming cutoff scale of dark matter at least as well as current strong-lensing and Milky Way satellite-count limits, with the advantage of being a purely local, gravitational probe whose systematics are independent of those of the other methods. This is especially timely given the kinematic data now becoming available from Gaia~\cite{Tavangar_2025}, DESI~\cite{Jarvis_2026}, and dedicated upcoming facilities such as the VIA Project~\cite{VIA_project}. 

Several directions follow naturally from this work. The most immediate is the application of the framework to actual GD-1 data. The harder and more interesting open problems are incorporating baryonic perturbers and handling the rare massive encounters that fall outside the diffusion regime, possibly through a hybrid scheme; we leave these for future work.

\section{Acknowledgments}

We thank Sten Delos for useful discussions and for comments on the draft. NAM thanks Adri Duivenvoorden for suggesting the use of the NUFFT early in the work and for helpful discussions.

Our analyses are performed on the \href{https://docs.mpcdf.mpg.de/doc/computing/clusters/systems/Astrophysics/MPA-FREYA.html}{\texttt{FREYA}} cluster, maintained by the Max Planck Computing \& Data Facility.

\textit{Software:} We ackowledge the use of the following softwares \jax \cite{jax2018github}, \galax \cite{galax}, \texttt{diffrax} \cite{diffrax}, \texttt{unxt} \cite{unxt}, \texttt{jax-finufft} at {\url{https://github.com/flatironinstitute/jax-finufft}} based on \cite{FINUFFT}.

\bibliography{references}

\appendix

\section{Runtime comparison} \label{app:runtime}

\begin{figure}[t]
    \centering
    \includegraphics[width=\linewidth]{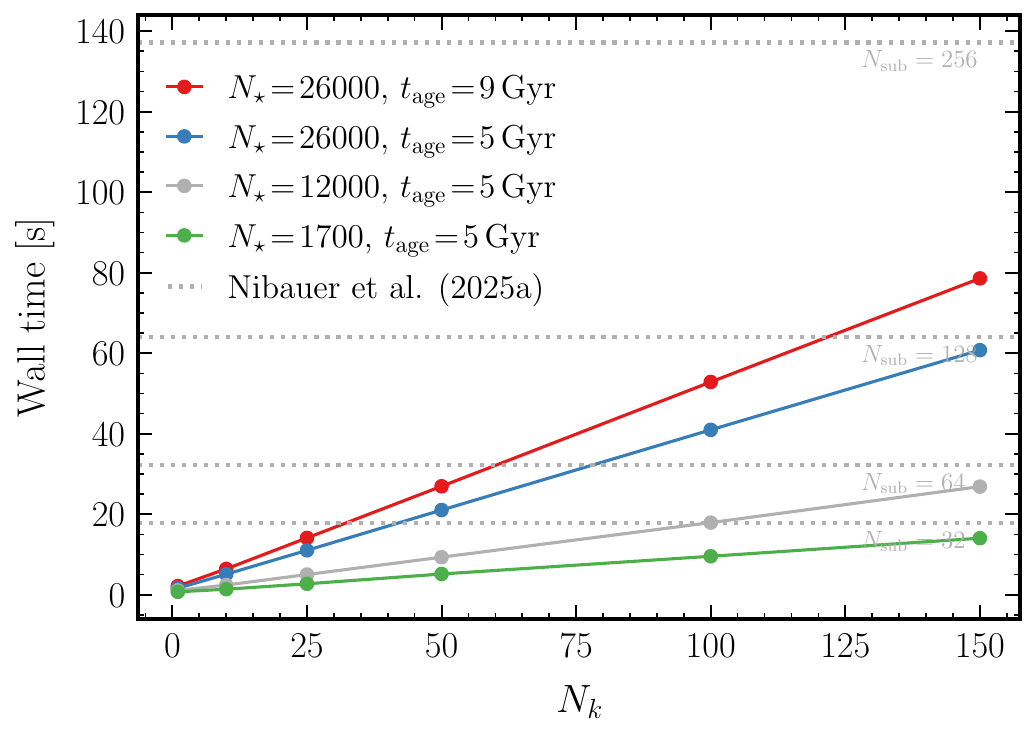}
    \caption{Post-compilation wall-clock time on a single NVIDIA A100 GPU as a function of the number of kick times $N_k$, for four stream configurations. The one-off \texttt{jit} compilation overhead, not shown, amounts to $\sim$  $20$~s regardless of $N_k$. All three configurations use the same potential and substructure setup as in Sec.~\ref{subsec:comparison}, with $N_\delta=1$ substructure field and grid resolution $\eta = 0.05$~kpc. Horizontal dotted lines indicate wall-clock times from Ref.~\cite{Nibauer:2024uue} (their Appendix~C, 1~GPU) for $N_{\rm sub} = 32$, $64$, $128$, and $256$ individually resolved subhalo encounters in a setup similar to our third configuration (in gray in the plot). For reference, results presented in this paper used $N_k$ in the range $30-70$.
    Note that $N_k$ and $N_{\rm sub}$ are not directly comparable (see text); the comparison is shown for order-of-magnitude orientation only.}
    \label{fig:runtime}
\end{figure}

Fig.~\ref{fig:runtime} shows the wall-clock time of our simulator as a function of the number of kick times $N_k$, measured on a single NVIDIA A100 GPU with 40GB of memory. All our timings use double-precision (\texttt{float64}) arithmetic. We benchmark three configurations: our fiducial production setup used for the GD-1 analysis ($N_\star = 26000$, $t_{\rm age} = 9$~Gyr) in Sec.~\ref{subsec:comparison}, a reduced-age variant ($N_\star = 26000$, $t_{\rm age} = 5$~Gyr), and a lighter configuration ($N_\star = 12000$, $t_{\rm age} = 5$~Gyr) chosen to match the particle count and stream age used in Ref.~\cite{Nibauer:2024uue} for direct comparison, and the setup used in the forecasts (Sec.~\ref{sec:forecasts}) for the GD-1 analysis ($N_\star = 1700$, $t_{\rm age} = 5$~Gyr) . In all cases we adopt $N_\delta = 1$, $\eta = 0.05$~kpc, and the same potential and substructure setup as in Sec.~\ref{subsec:comparison}.

The runtime scales \emph{linearly} with $N_k$ in all configurations. The timings shown in Fig.~\ref{fig:runtime} correspond to the post-compilation runtime, i.e., the wall-clock time of the already \texttt{jit}-compiled simulation. The one-off \texttt{jit} compilation overhead, not shown in the figure, amounts to $\sim$ $20$~s regardless of $N_k$. This linear scaling is expected: each kick step consists of a fixed  amount of work (one NUFFT evaluation and one orbital integration step via \texttt{galax}) and these steps are executed sequentially by \texttt{jax.lax.scan} with no dependence on the results of  previous kicks beyond the updated particle positions. The per-star computations within each step are parallelised via \texttt{vmap} and executed on GPU. As noted in Sec.~\ref{subsec:simulations}, \texttt{jit} compilation requires that both the kick times and  stripping times are fixed at compile time; our assumption of uniformly spaced times satisfies this requirement and enables the efficient \texttt{scan}-based loop. 

Comparing the first three configurations, reducing $t_{\rm age}$ from  9 to 5~Gyr at fixed $N_\star = 26000$ decreases the runtime  by roughly $40\%$, reflecting the shorter orbital integration  between consecutive kick times. More than halving the number of particles to $N_\star = 12000$ at fixed $t_{\rm age} = 5$~Gyr yields a further reduction of $\sim 30\%$, confirming that both the particle count and the integration time contribute comparably to the per-step cost.

For context, we compare our runtimes to those reported in Ref.~\cite[][App.~C]{Nibauer:2024uue} for their perturbative stream model, which resolves the dynamical effect of each subhalo encounter individually using Hamiltonian perturbation theory. The horizontal dotted lines in Fig.~\ref{fig:runtime} indicate the wall-clock times measured in our digitized version of Ref.~\cite[Fig.~19,][]{Nibauer:2024uue} for $N_{\rm sub} = 32$, $64$, $128$, and $256$ individually resolved subhalo encounters. We stress that $N_k$ and $N_{\rm sub}$ are \emph{not} directly comparable quantities: $N_k$ counts the number of times we sample the full statistical substructure field in the diffusion regime, while $N_{\rm sub}$ counts the number of discrete subhalo encounters resolved individually. The two approaches are therefore complementary rather than competing: Ref.~\cite{Nibauer:2024uue} is optimised for the regime of a small number of massive, individually resolvable encounters, whereas our diffusion-regime framework is designed to capture the collective statistical effect of an arbitrarily large population of low-mass subhalos that would be prohibitively expensive to resolve individually. 

\vspace{20pt}
\section{Convergence tests} \label{app:convergence}

\begin{figure*}
    \centering
    \includegraphics[width=0.9\linewidth]{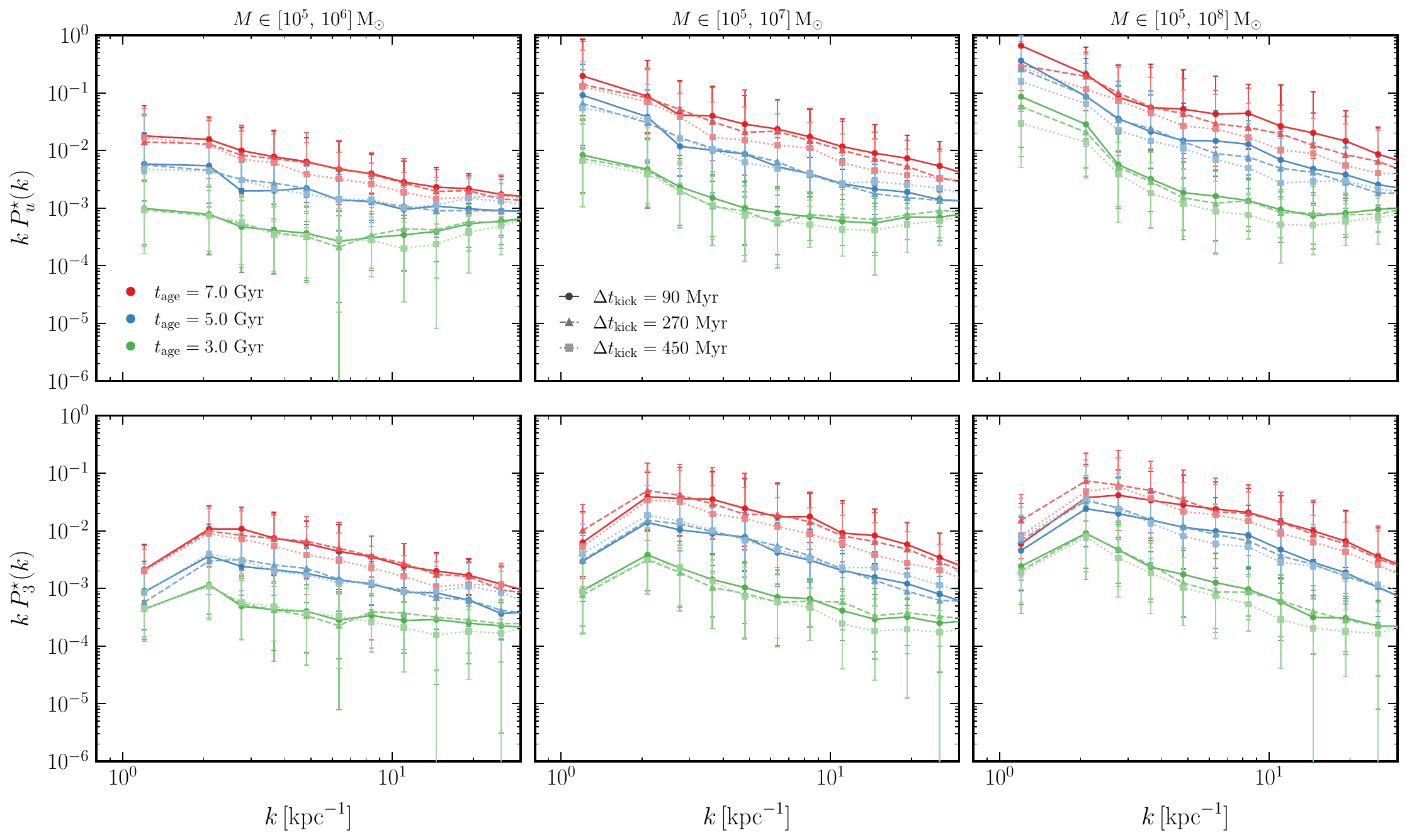}
    \caption{Convergence of the stream density power spectrum with respect to the kick interval $\Delta t \in \{90, 270, 450\}$~Myr, with all other parameters fixed at their fiducial values. Layout as in Fig.~\ref{fig:comparison}.}
    \label{fig:convergence_kick}
\end{figure*}

\begin{figure*}
    \centering
    \includegraphics[width=0.9\linewidth]{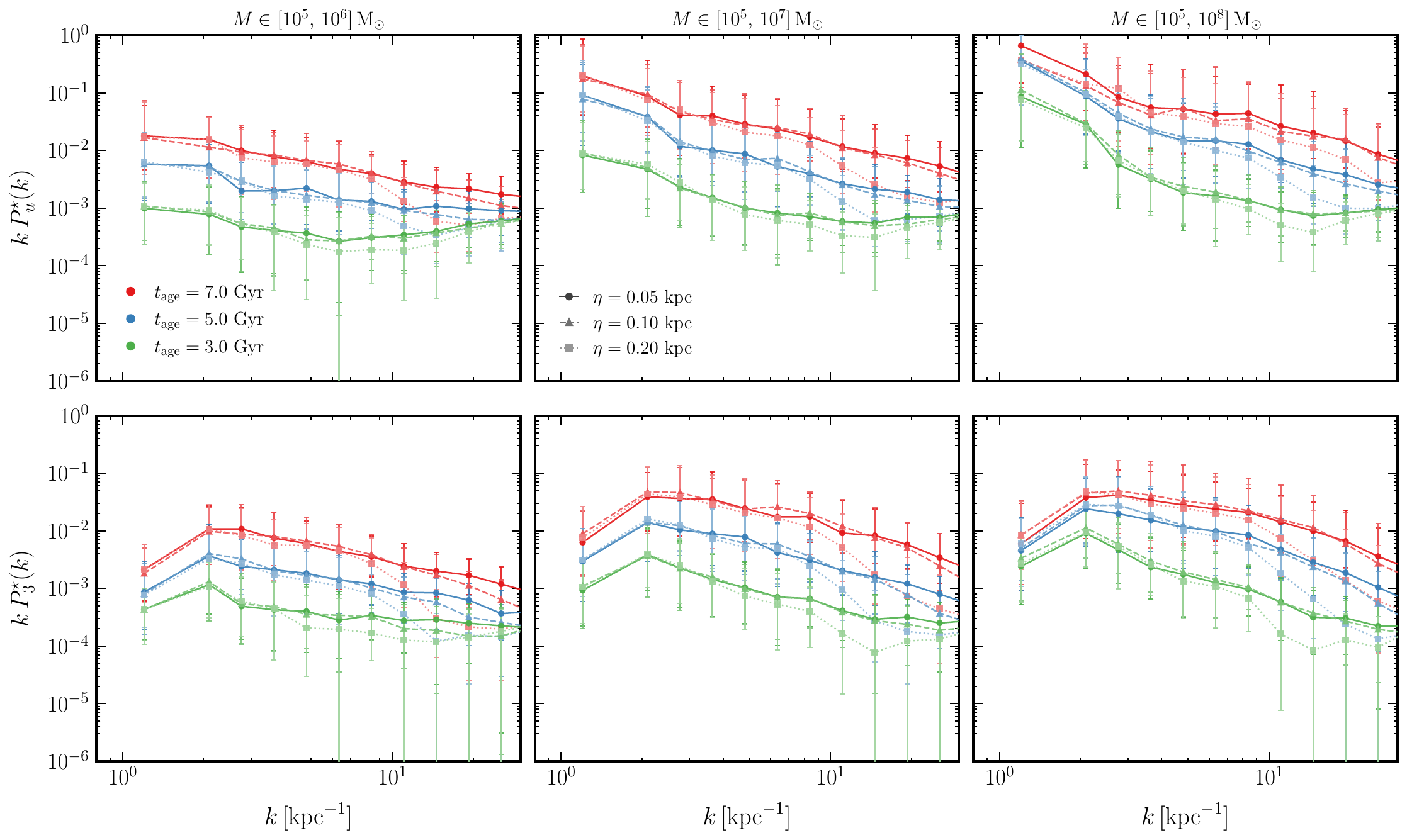}
    \caption{Same as Fig.~\ref{fig:convergence_kick}, but varying the grid resolution $\eta \in \{0.05, 0.10, 0.20\}$~kpc.}
    \label{fig:convergence_resolution}
\end{figure*}

\begin{figure*}
    \centering
    \includegraphics[width=0.9\linewidth]{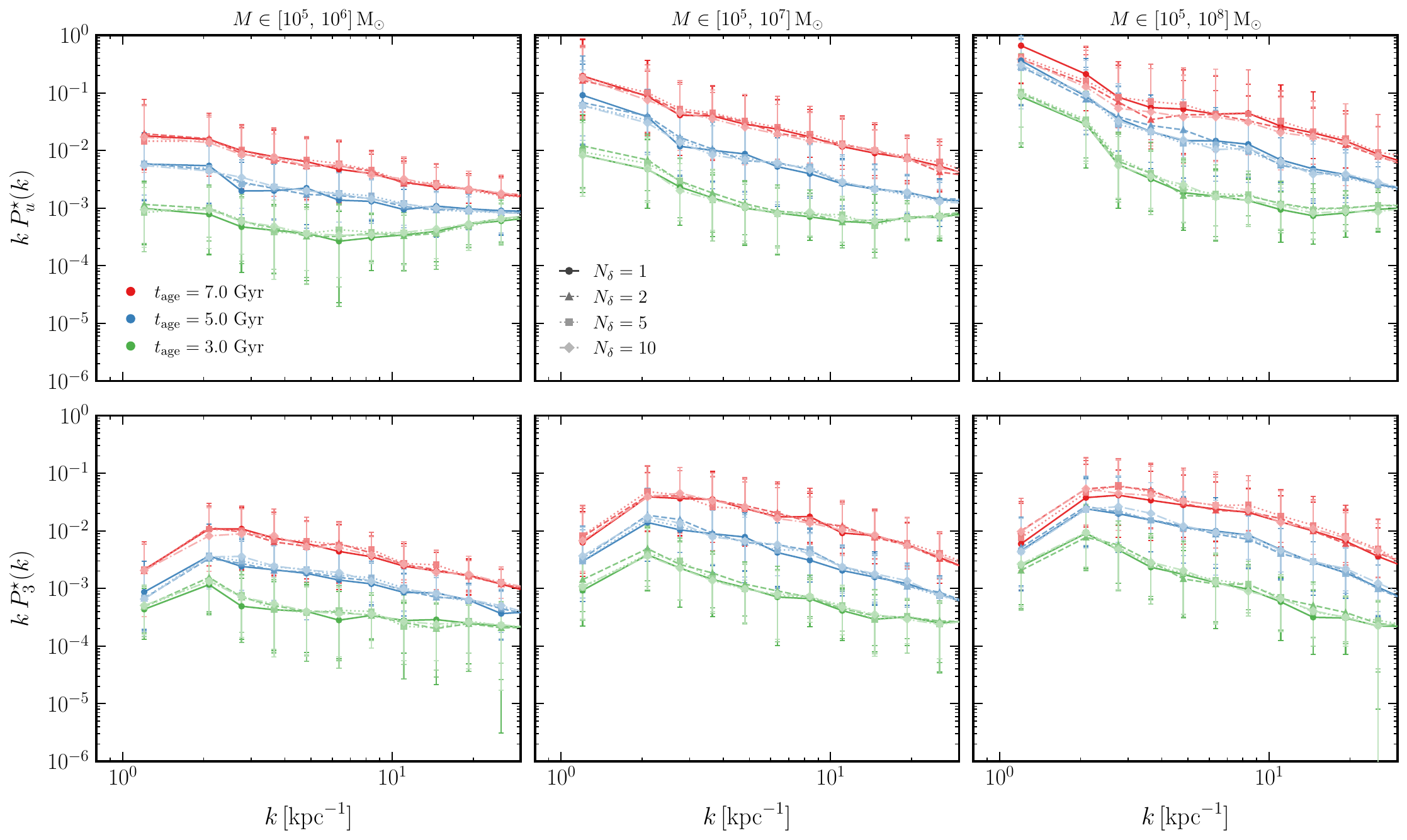}
    \caption{Same as Fig.~\ref{fig:convergence_kick}, but varying the number of substructure fields per kick $N_\delta \in \{1, 2, 5, 10\}$.}
    \label{fig:convergence_ndelta}
\end{figure*}

We test the convergence of the stream density power spectra with respect to the three main simulation parameters: the kick interval $\Delta t$, the grid resolution $\eta$, and the number of substructure fields per kick $N_\delta$. In each test, we vary one parameter while keeping the others fixed at their fiducial values ($\Delta t = 90$~Myr, $\eta = 0.05$~kpc, $N_\delta = 1$), and compare the resulting density power spectra across all mass ranges and stream ages considered in Sec.~\ref{subsubsec:realistic}. The layout of each figure mirrors that of Fig.~\ref{fig:comparison}: rows correspond to the two density contrast estimators $\delta^*_u$ and $\delta^*_3$, columns to mass ranges $M \in [10^5, 10^6]$, $[10^5, 10^7]$, and $[10^5, 10^8]\,\mathrm{M}_\odot$, and colors to stream ages $t_{\rm age} = 7$, $5$, and $3$~Gyr.

Figs.~\ref{fig:convergence_kick}, \ref{fig:convergence_resolution}, and~\ref{fig:convergence_ndelta} show the results for $\Delta t \in \{90, 270, 450\}$~Myr, $\eta \in \{0.05, 0.10, 0.20\}$~kpc, and $N_\delta \in \{1, 2, 5, 10\}$, respectively. In all three cases, the power spectra are consistent within error bars across all wavenumber bins, mass ranges, and stream ages, confirming that the fiducial parameter choices are well converged for the substructure scenarios and summary statistics considered here.

We note that convergence in simulation parameters is necessarily summary-statistic dependent: the choices validated here for the density power spectrum may require re-evaluation if different observables or higher-order statistics are used as the basis for inference.
We verified that the auto- and cross-spectra of the full kinematic data vector used in the forecasts (Sec.~\ref{subsec:forecast_observables}) are also well converged with these parameter choices.

\end{document}